\documentstyle[epsfig,12pt]{article}  
\newcommand{\bce}{\begin{center}}
\newcommand{\ece}{\end{center}}
\newcommand{\beq}{\begin{equation}}
\newcommand{\eeq}{\end{equation}}
\newcommand{\bea}{\vspace{0.25cm}\begin{eqnarray}}
\newcommand{\eea}{\end{eqnarray}}

\newcommand{\br}{{\bf
r}}

\newcommand{\ba}{\begin{array}}
\newcommand{\ea}{\end{array}}

\newcommand{\r}{\mbox{{\boldmath
$\rho$}}}
\newcommand{\ta}{\mbox{{\boldmath
$\tau$}}}

\newcommand{\bfa}{\mbox{{\boldmath
$\alpha$}}}

\newcommand{\vb}{\mbox{{\bf
v}}}
\newcommand{\qb}{\mbox{{\bf
q}}}

\newcommand{\bb}{\mbox{{\bf
b}}}
\newcommand{\rb}{\mbox{{\bf
r}}}
\newcommand{\eb}{\mbox{{\bf
e}}}

\newcommand{\doublespace}{
\renewcommand{\baselinestretch}{1.6}\large\normalsize}

\def\lsim{\mathrel{\rlap{\lower4pt\hbox{\hskip1pt$\sim$}}
    \raise1pt\hbox{$<$}}}         %less than or approx. symbol
\def\gsim{\mathrel{\rlap{\lower4pt\hbox{\hskip1pt$\sim$}}
    \raise1pt\hbox{$>$}}}         %greater than or approx. symbol

    \def\Pom{{\bf
    I\!P}}
    \def\beq{\begin{equation}}
    \def\endeq{\end{equation}}
    \def\bea{\begin{eqnarray}}
    \def\arr{\begin{eqnarray}}
    \def\eea{\end{eqnarray}}

\def\q2{$Q^{2}$}
\def\s2{2$S$}

\begin{document}
\doublespace
\thispagestyle{empty}
\vspace*{-2cm}
\begin{flushright}
\bf
MPI-H-V44-1997\\
%\vspace{-1.cm}
%December
%1997\\
\end{flushright}
\bigskip

\begin{center}

  {\Large\bf
  
  Light-cone path integral approach to
  the
  Landau--Pomeranchuk--Migdal
  effect
  }
  
  \vspace{.5
  cm}
  
  {\large B.G. Zakharov
  }
  \bigskip
%  \bigskip

  {\it
  Max--Planck Institut f\"ur Kernphysik, Postfach
  103980\\
  69029 Heidelberg,
  Germany\medskip\\
  L. D. Landau Institute for Theoretical
  Physics,
  GSP-1, 117940,\\ ul. Kosygina 2, 117334 Moscow,
  Russia
  \medskip\\}
  \vspace{.5cm}

  {\bf\large
  Abstract}
\end{center}
A new rigorous light-cone path integral approach
to
the Landau-Pomeranchuk-Migdal effect in QED and
QCD is discussed.
The rate of photon (gluon)
radiation
by an electron (quark) in a medium is expressed
through
the Green's function of a two-dimensional
Schr\"odinger
equation with an imaginary potential. In QED this
potential
is proportional to the dipole cross section for
scattering
of $e^{+}e^{-}$ pair off an atom, in QCD it
is
proportional to the cross section of
interaction
of the color singlet quark-antiquark-gluon
system
with a medium
constituent.
In QED our predictions agree
well
with the photon spectrum measured recently at SLAC
for
25 GeV
electrons.
In QCD for a sufficiently energetic quark produced inside a
medium
we predict the radiative energy loss $\Delta E_{q}\propto
L^{2}$,
where $L$ is the distance passed by the quark in
the
medium. It has a weak dependence on the initial quark
energy
$E_{q}$. The $L^{2}$ dependence transforms into $L^{1}$ as the quark
energy
decreases.
We also give new formulas
for
nuclear shadowing in hard
reactions.

\vspace*{3cm}

\begin{center}
{E-mail: bgz@landau.ac.ru}
\end{center}

\noindent
\newpage
%-------------------------------------------------------------
\section{Introduction}
%-------------------------------------------------------------
In 1953 Landau and Pomeranchuk \cite{LP}
predicted
within classical
electrodynamics
that
multiple
scattering
can considerably suppress bremsstrahlung of
high
energy charged particles in
a medium.
In the high energy limit they obtained
the
photon radiation rate
$\propto
1/\sqrt{\omega}$ ($\omega$ is the photon frequency),
which
differs
drastically
from the spectrum for an isolated atom $\propto
1/\omega$.
Later, this result was confirmed by Migdal \cite{Migdal},
who
developed a quantum-mechanical theory of this
phenomenon.
The physical mechanism behind the suppression of the
radiation
rate in a medium is the loss of coherence for photon emission
from
different parts of the
charged
particle trajectory at the scale of the photon formation
length.

Since the studies by Landau and Pomeranchuk \cite{LP} and
Migdal
\cite{Migdal},
the suppression of radiation processes in
medium,
called in the current literature the Landau-Pomeranchuk-Migdal
(LPM)
effect, has been studied in many theoretical papers
[3-21].
However, only recently the
first
quantitative measurement of the LPM effect
for
high energy electrons was performed at SLAC
\cite{SL1}.
Altogether, this experiment corroborated LPM
suppression
of
bremsstrahlung.
Unfortunately, contamination of the SLAC
data
by multiphoton emission makes it difficult to perform
an
accurate comparison with the theoretical predictions in the
entire
range of photon
energies. Nonetheless,
the experimental spectrum for a thin gold
target
at photon energies from 5 MeV up to 500 MeV
for
25 GeV electron beam, where the multiphoton emission gives a
small
contribution, agrees with predictions of
Ref. \cite{LPM2}.
The calculations of Ref. \cite{LPM2} were
carried
out within a new light-cone path integral approach to the LPM
effect
which we
developed
in Ref. \cite{LPM1}. In Ref. \cite{LPM2}
the
radiation rate was calculated including inelastic processes
and
treating rigorously the Coulomb
effects.
In all previous analyses the inelastic processes were
neglected,
and the Coulomb effects were treated in the leading-log
approximation.
This approximation works well in the limit
of
strong LPM suppression in an infinite
medium. However,
it is not good for real
situations
because of the uncertainty in the value of
the
Coulomb
logarithm.

The approach of Ref. \cite{LPM1} is also applicable in
QCD.
Analysis of the LPM effect in
QCD
is of great importance for understanding
the
longitudinal energy flow in soft and hard
hadron-nucleus
collisions and the energy loss of a fast quark
produced
in deep inelastic scattering on nuclear
target.
It becomes especially
interesting
in connection with the forthcoming
experiments
on high energy $AA$-collisions at RHIC and LHC,
where
the formation of quark-gluon plasma (QGP) is
expected.
The energy loss of high-$p_{\perp}$ jets produced at
the initial
stage of
$AA$-collision
may be an important potential probe for formation of
QGP.
The first attempt to estimate the radiative quark energy loss
$\Delta
E_{q}$ in QGP was made by Gyulassy and Wang \cite{GW1}. They
modelled
QGP by a system of static scattering centers described by
the
Debye screened Coulomb potential $\propto \exp(-r\mu_{D})/r$,
where
$\mu_{D}$ is the color screening
mass. The
authors studied emission of soft
gluons
in the region of
small
transverse momenta $k_{\perp}<<\mu_{D}$, which, however, gives
a
negligible contribution to $\Delta
E_{q}$.
Analysis of soft gluon radiation without
a
restriction
on
the gluon transverse momentum in the limit of
strong
LPM
suppression
within the Gyulassy-Wang (GW) model was performed in
Refs.
\cite{B1,B3}.
A similar analysis for cold nuclear matter was given
in
Ref. \cite{Levin}.
However, the authors of Refs. \cite{B1,Levin,B3} used
some
unjustified approximations. For
instance,
the quark-gluon system emerging after
gluon
emission was treated as a pointlike color triplet
object.
\footnote{
Note that, after submission of the present paper in November 1997 to
{\sl Phys. At. Nucl.},
R.Baier, Yu.L.Dokshitzer, A.H.Mueller and
D.Schiff ({\sl hep-ph}/9804212) had reanalyzed the
induced gluon radiation without using the approximation of
the pointlike $qg$ system.}
A rigorous quantum treatment of
the
induced gluon radiation was given for the first time in
Ref.
\cite{LPM1} (see also
\cite{LPM3}).

In the present paper we discuss the LPM effect in QED and
QCD
within the approach of Ref. \cite{LPM1}. We give special attention 
to
technical details omitted
in
our previous short publications. We also consider from
the
viewpoint of LPM suppression nuclear shadowing
in
hard reactions. New formulas
for
shadowing that take into account the parton
transverse
motion are derived.

The presentation is organized as
follows.
Section 2 is devoted to the LPM effect in
QED.
Using the unitarity of scattering
matrix
we express the cross section of photon emission
through
radiative correction to the
transverse
electron propagator. It is
calculated
within time-ordered perturbation theory (PT) in
coordinate
representation.
The radiation rate is expressed through the Green's
function
of a two-dimensional Schr\"odinger equation with an
imaginary
potential which is proportional to the dipole cross
section
for scattering of $e^{+}e^{-}\gamma$ state off an
atom.
We demonstrate that the cross section of photon emission
can
be written in the form analogous to the Glauber amplitude
for
elastic hadron-nucleus scattering. This
representation
allows one to view
LPM
suppression as an absorption effect for $e^{+}e^{-}\gamma$
system.
We compare the theoretical
predictions
with the data of the SLAC experiment
\cite{SL1}.
In section 3 we discuss the LPM effect in
QCD.
As in QED, the radiation rate
is
expressed through the Green's function of a
two-dimensional
Schr\"odinger equation. The corresponding
imaginary
potential is proportional to the total
cross
section for a three-body quark-antiquark-gluon
state.
We evaluate the quark energy loss for cold
nuclear
matter and QGP using the oscillator parametrization for the
imaginary
potential.
For a high energy
quark
incident on a nucleus we
predict
$\Delta E_{q}\sim 0.1
E_{q}(L/10\,\mbox{fm}$).
For a fast quark produced
inside
a medium we obtain $\Delta E_{q}\propto
L^{2}$
($L$ is the quark path length in the medium) while at sufficiently small
energy
$\Delta E_{q}\propto
L$.
In section 4 we discuss the LPM effect for hard reactions on
nuclear
targets. We demonstrate that our approach to the LPM
effect
can be used for an accurate evaluation of nuclear
shadowing.
In section 5 we summarize our
results.

%-------------------------------------------------------------------
\section{The LPM effect in QED}
%-------------------------------------------------------------------
\subsection{ General expression for the radiation rate}
%-------------------------------------------------------------------
We begin with the LPM effect for bremsstrahlung of a fast
electron.
We consider an electron
incident
on an amorphous target of a finite
thickness.
Multiphoton emission will be
neglected.
The probability
of
photon
emission, $P_{\gamma}$, is connected with the probability $P_{e}$
to
detect in
the final
state one electron by the unitarity
relation:
\beq
P_{e}+P_{\gamma}=1\,.
\label{eq:2.10}
\eeq
In the absence of interaction of the electron with
the
quantum photon field, we have  $P_{e}=1$. Consequently,
Eq. (\ref{eq:2.10})
can be rewritten
as
\beq
P_{\gamma}=-\left(\delta P_{e}-\delta
P_{e}^{vac}\right)\,,
\label{eq:2.20}
\eeq
where $\delta P_{e}$ is
the
radiative correction of order $\alpha$ ($\alpha=1/137$) to
$P_{e}$.
On the right-hand side of Eq. (\ref{eq:2.20}) we
subtracted
the vacuum term,
which
takes into account the renormalization of the electron wave
function
for initial and final electron
states.

We will evaluate $\delta P_{e}$ in time-ordered
PT
\cite{Wein,BKS}. The corresponding matrix element is
generated
by
transitions
$e\rightarrow e'\gamma\rightarrow
e$.
In a medium the electron transverse
momentum
is not
conserved.
For this reason it is convenient to use the
coordinate
representation of time-ordered PT in
which
the transitions $e\rightarrow e'\gamma\rightarrow e$ will
reveal
themselves through the radiative correction to the electron
wave
function.
Let us first
consider
the wave function of a
fast
electron neglecting interaction with the quantum photon
field.
In the vacuum the
radiative-correction-free wave function of a relativistic
electron
with longitudinal momentum $p_{z}>>m_{e}$ ($m_{e}$ is
the
electron mass) can be written
as
\beq
\psi(t,\rb)=\exp[-ip_{z}(t-z)]\phi(t,\r)\,,
\label{eq:2.30}
\eeq
where $\rb=(z,\r)$, and the time-dependence of
the
transverse wave function $\phi(t,\r)$ is governed
by
the
two-dimensional
Schr\"odinger
equation
\beq
i\frac{\partial\phi(t,\r)}{\partial
t}=
{H}\phi(t,\r)\,,
\label{eq:2.40}
\eeq
with the
Hamiltonian
\beq
{H}=\frac{({\qb}^{2}+m^{2}_{e})}{2\mu_{e}}\,.
\label{eq:2.50}
\eeq
Here $\qb$ is the operator of
transverse
momentum, and the Schr\"odinger
mass is
$\mu_{e}=p_{z}$. Eqs. (\ref{eq:2.30}), (\ref{eq:2.40}) hold for
each
helicity
state. At high energy the electron propagates nearly along
the
light-cone $t-z=\mbox{const}$, and in Eq. (\ref{eq:2.40})
the
variable $t$ can be viewed as the longitudinal coordinate
$z$.
For this reason, we will henceforth regard the transverse
wave
function $\phi$ as a function of $z$ and
$\r$.
Eq. (\ref{eq:2.40}) allows one to write the following
relation,
connecting
$\phi(z,\r)$
at planes $z=z_{1}$
and
$z=z_{2}$,
\beq
\phi(z_{2},\r_{2})=\int
d\r_{1}K_{e}(\r_{2},z_{2}|\r_{1},z_{1})
\phi(z_{1},\r_{1})\,\,,
\label{eq:2.60}
\eeq
where
\beq
K_{e}(\r_{2},z_{2}|\r_{1},z_{1})=
\frac{\mu_{e}}{2\pi i\Delta
z}
\exp\left[\frac{i\mu_{e}(\r_{2}-\r_{1})^{2}}
{2\Delta
z}
-\frac{i m_{e}^{2}\Delta
z}{2\mu_{e}}\right]
\label{eq:2.70}
\eeq
is the Green's function of
the
two-dimensional Hamiltonian (\ref{eq:2.50}), with
$\Delta
z=z_{2}-z_{1}$.
At high energies spin effects in interaction of an
electron
with an atom vanish, and equations
analogous
to (\ref{eq:2.30}) and (\ref{eq:2.60}) hold for propagation of
an
electron
through the medium as well. The corresponding
propagator
can be
written
in the Feynman path integral
form
\beq
K_{e}(\r_{2},z_{2}|\r_{1},z_{1})=
\int {\cal
D}\r
\exp
\left\{
i\int
dz
\left[
\frac{\mu_{e} \dot{\r}^{2}}{2}+e\,
U(\r,z)
\right]-\frac{i m_{e}^{2}\Delta
z}{2\mu_{e}}
\right\}\,\,,
\label{eq:2.80}
\eeq
where $\dot{\r}=d\r/dz$, and $U(\r,z)$ is the potential of the
medium.

The photon wave function can also be written in the
form
(\ref{eq:2.30}). Using the representation (\ref{eq:2.30}) for
the
electron
and
photon wave functions, we 
can
obtain for the radiative correction to the
transverse
electron
propagator
associated with $
e'\gamma$
intermediate
state
\bea
\delta
K_{e}(\r_{2},z_{2}|\r_{1},z_{1})=-
\int\limits_{0}^{1}
dx
\int\limits_{z_{1}}^{z_{2}} d
\xi_{1}\int\limits_{\xi_{1}}^{z_{2}}
d
\xi_{2}
\int d \ta_{1}d
\ta_{2}
g(\xi_{1},\xi_{2},x)\nonumber\\
\times
K_{e}(\r_{2},z_{2}|\ta_{2},\xi_{2})
K_{e'}(\ta_{2},\xi_{2}|\ta_{1},\xi_{1})
K_{\gamma}(\ta_{2},\xi_{2}|\ta_{1},\xi_{1})
K_{e}(\ta_{1},\xi_{1}|\r_{1},z_{1})\,.
\label{eq:2.90}
\eea
Here
the indices $e'$ and $\gamma$ label
the
electron and photon
propagators
for the intermediate $e'\gamma$
state.
The Schr\"odinger masses that appear in the Green's functions
$K_{e'}$
and $K_{\gamma}$
are
$\mu_{e'}=(1-x)\mu_{e}$ and $\mu_{\gamma}=x\mu_{e}$,
where
$x$ is the light-cone fractional momentum of
the
photon.
The vertex operator $g(\xi_{1},\xi_{2},x)$,
including
all spin effects associated with
transitions
$e\rightarrow e'\gamma\rightarrow
e$,
is given
by
\beq
g(\xi_{1},\xi_{2},x)=
\frac{\alpha[4-4x+2x^{2}]}{4x}\,
\vb(\xi_{2})\cdot\vb(\xi_{1})+
\frac{\alpha
m_{e}^{2}x}{2\mu_{e'}^{2}}\,\,,
\label{eq:2.100}
\eeq
where
$$
\vb(\xi_{i})={\vb}_{\gamma}(\xi_{i})
-{\vb}_{e'}(\xi_{i})\,,
$$
$\vb_{\gamma}$ and $\vb_{e'}$ are
the
transverse velocity operators, which act on
the
corresponding propagators in
Eq.~(\ref{eq:2.90}).
Two terms on the right-hand side of
Eq. (\ref{eq:2.100})
correspond to the $e\rightarrow e'\gamma$
transitions
conserving and changing the electron
helicity.

Now we have all ingredients necessary for calculating 
the
radiation
rate.
We consider a target with a density
$n(z)$
independent of the impact parameter, and assume
that
$n(z)$ vanishes as $|z|\rightarrow
\infty$.
In terms of the transverse Green's
function
$K_{e}$
and
$\delta K_{e}$ the radiative correction $\delta
P_{e}$
can be written
as
\beq
\delta P_{e}=2\mbox{Re}\int d
\r_{1}
d
\r_{1}'
d \r_{2}
\phi(z_{1},\r_{1})\phi^{*}(z_{1},\r_{1}')
\langle
\delta
K_{e}(\r_{2},z_{2}|\r_{1},z_{1})
K^{*}_{e}(\r_{2},z_{2}|\r_{1}',z_{1})
\rangle
\,,
\label{eq:2.110}
\eeq
where $\langle\,...\,\rangle$ means averaging
over
the states of the
target,
and the
points
$z_{1}$ and $z_{2}$ are assumed to be at
large
distances before and after the target, respectively. The
initial
electron wave function is normalized by the condition
\beq
\int d\r
|\phi(z_{1},\r)|^{2}=1\,\,.
\label{eq:2.111}
\eeq

For a high energy electron one can neglect the longitudinal
momentum
transfer associated with the interaction with a medium
potential.
For this reason the unitarity relation (\ref{eq:2.20}) is
also
valid in the differential form in the light-cone variable
$x$.
Then, using Eqs. (\ref{eq:2.20}), (\ref{eq:2.90})
and
(\ref{eq:2.110}),
we
can obtain for the radiation
rate
[we suppress the vacuum term, which will be recovered in the
final
formula
(\ref{eq:2.290})]
\bea
\frac{d P_{\gamma}}{d
x}=2\mbox{Re}
\int\limits_{z_{1}}^{z_{2}} d
\xi_{1}
\int\limits_{\xi_{1}}^{z_{2}}d
\xi_{2}\int
d \r_{1} d\r_{1}'
d\ta_{1}d\ta_{1}'
d\ta_{2}d\ta_{2}'d\r_{2}
\phi(z_{1},\r_{1})\phi^{*}(z_{1},\r_{1}')
\nonumber\\
\times
g(\xi_{1},\xi_{2},x)
S(\r_{2},\r_{2},z_{2}|\ta_{2},\ta_{2}',\xi_{2})
M(\ta_{2},\ta_{2}',\xi_{2}|\ta_{1},\ta_{1}',\xi_{1})
S(\ta_{1},\ta_{1}',\xi_{1}|\r_{1},\r_{1}',z_{1})
\,,
\label{eq:2.120}
\eea
where
\beq
S(\r_{2},\r_{2}',\xi_{2}|\r_{1},\r_{1}',\xi_{1})
=\langle
K_{e}(\r_{2},\xi_{2}|\r_{1},\xi_{1})
K^{*}_{e}(\r_{2}',\xi_{2}|\r_{1}',\xi_{1})
\rangle
\label{eq:2.130}
\eeq
is the evolution operator for the electron density
matrix
in the absence of interaction with the photon
field,
and
\beq
M(\r_{2},\r_{2}',\xi_{2}|\r_{1},\r_{1}',\xi_{1})=
\langle
K_{e'}(\r_{2},\xi_{2}|\r_{1},\xi_{1})
K_{\gamma}(\r_{2},\xi_{2}|\r_{1},\xi_{1})
K^{*}_{e}(\r_{2}',\xi_{2}|\r_{1}',\xi_{1})
\rangle\,.
\label{eq:2.140}
\eeq
In deriving Eq. (\ref{eq:2.120}) we used
the
convolution
relation
\beq
K_{e}(\r_{2},z_{2}|\r_{1},z_{1})=
\int
d\ta_{1}d\ta_{2}
K_{e}(\r_{2},z_{2}|\ta_{2},\xi_{2})
K_{e}(\ta_{2},\xi_{2}|\ta_{1},\xi_{1})
K_{e}(\ta_{1},\xi_{1}|\r_{1},z_{1})\,.
\label{eq:2.141}
\eeq

Using the path integral representation for
the
transverse Green's functions, we can rewrite
Eqs. (\ref{eq:2.130}),
(\ref{eq:2.140}) in the
form
\beq
S(\r_{2},\r_{2}',\xi_{2}|\r_{1},\r_{1}',\xi_{1})
=\int{\cal D}\r_{e}{\cal
D}\r_{e}'
\exp\left[\frac{i\mu_{e}}{2}\int
d\xi(\dot{\r}_{e}^{2}
-\dot{\r}_{e}'^{2})\right]
\Phi(\{\r_{e}\},\{\r_{e}'\})\,,
\label{eq:2.150}
\eeq
\bea
M(\r_{2},\r_{2}',\xi_{2}|\r_{1},\r_{1}',\xi_{1})=
\int {\cal D}\r_{e'}{\cal D}\r_{\gamma}{\cal
D}\r_{e}
\exp\left\{
\frac{i}{2}\int d
\xi
(\mu_{e'}\dot{\r}_{e'}^{2}+\mu_{\gamma}\dot{\r}_{\gamma}^{2}
-\mu_{e}\dot{\r}_{e}^{2})\right.\nonumber\\
\left.
-\frac{i(\xi_{2}-\xi_{1})}{L_{f}}\right\}
\Phi(\{\r_{e'}\},\{\r_{e}\})\,\,.
\label{eq:2.160}
\eea
Here we introduced the photon formation
length
\beq
L_{f}=2\mu_{e}\left[\frac{m_{e}^{2}}{1-x}-m_{e}^{2}\right]^{-1}=
\frac{2E_{e}(1-x)}{m_{e}^{ 2 }x}
\,,
\label{eq:2.161}
\eeq
which appears in Eq. (\ref{eq:2.160}) owing to the difference
between
the phases of the wave functions of $e$ and $e'$
states.
Notice that the value of $L_{f}$, emerging
in
deriving Eq. (\ref{eq:2.160}), agrees with the
estimate
based on the uncertainty relation $\Delta E\Delta t\sim
1$.
In the following we will assume that $L_{f}$ is much larger
than
the atomic
size.
The boundary conditions for trajectories in
Eq.~(\ref{eq:2.150})
are
$\r_{e}(\xi_{1,2})=\r_{1,2}$,
$\r_{e}'(\xi_{1,2})=\r_{1,2}'$, and in
Eq.~(\ref{eq:2.160})
$\r_{e',\gamma}(\xi_{1,2})=\r_{1,2}$,
$\r_{e}(\xi_{1,2})=\r_{1,2}'$.
The phase factor $\Phi$ in Eqs. (\ref{eq:2.150})
and
(\ref{eq:2.160}) that takes
into account interaction with a medium potential is given
by
\beq
\Phi(\{\r_{i}\},\{\r_{j}\})=
\left\langle\exp\left\{ie\int d
\xi
[U(\r_{i}(\xi),\xi)-U(\r_{j}(\xi),\xi)]\right\}\right\rangle\,.
\label{eq:2.170}
\eeq
Note that this phase factor can be viewed as the
one
for propagation through the medium of $e^{+}e^{-}$
system.
However, it should be borne in
mind
that the "positron" kinetic energy term in
Eqs. (\ref{eq:2.150}),
(\ref{eq:2.160}) is negative.

We will neglect the correlations in the positions of medium
atoms. In this case $\Phi(\{\r_{i}\},\{\r_{j}\})$
can
be written
as
\bea
\Phi(\{\r_{i}\},\{\r_{j}\})=
\left\{
1-\frac{1}{N}\int
d\xi
n(\xi)\right.\,\,\,\,\,\,\,\,\,\,\,\,\,\,\,\,\,\,\,\,\,
\nonumber\\
\times
\left.
\int
d{\bb}
\left\langle
1-\exp
\left\{ie\int d\xi'[\varphi
(\r_{i}(\xi')-{\bb},\xi'-\xi)-
\varphi
(\r_{j}(\xi')-{\bb},\xi'-\xi)]\right\}
\right\rangle_{a}
\right\}^{N}\,\,,
\label{eq:2.180}
\eea
where $\varphi(\br)$ is the atomic potential, $N$ is the
number
of the atoms in the
targets,
$\langle\,...\,\rangle_{a}$ denotes averaging
over
the states of the
atom.
After exponentiating Eq. (\ref{eq:2.180}) can be written in the
form
\beq
\Phi(\{\r_{i}\},\{\r_{j}\})=
\exp\left[-\frac{1}{2}
\int d\xi
n(\xi)\sigma(|\r_{i}(\xi)-\r_{j}(\xi)|)
\right]\,\,,
\label{eq:2.190}
\eeq
where
\beq
\sigma(|\r|)=
2\int
d{\bb}
\left\langle
1-\exp
\left\{ie\int d\xi[\varphi
(\r-{\bb},\xi)-
\varphi
({\bb},\xi)]\right\}\right\rangle_{a}
\label{eq:2.200}
\eeq
is the dipole cross
section
for scattering of $e^{+}e^{-}$ pair of the transverse
size
$\rho$ on the
atom.
In arriving at (\ref{eq:2.190}) we neglected the
variation
of $|\r_{i}(\xi)-\r_{j}(\xi)|$ at the longitudinal
scale
on the order of the atomic radius $a$. This is  a
good
approximation for
$L_{f}>>a$.

For an atomic
potential
$\varphi(r)=(Ze/4\pi
r)\exp(-r/a)$
($a\sim r_{B}Z^{-1/3}$, $r_{B}$ is the Bohr
radius)
$\sigma(\rho)$ in the Born approximation is given
by
\beq
\sigma(\rho)=8\pi(Z\alpha
a)^{2}
\left[1-\frac{\rho}{a}K_{1}\left(\frac{\rho}{a}\right)\right]\,,
\label{eq:2.201}
\eeq
where $K_{1}$ is the Bessel function. For $\rho\ll
a$,
which will be important in the problem under consideration, we have
$
\sigma(\rho)
\simeq
C(\rho)\rho^{2}\,,
$
where
\beq
C(\rho)=4\pi(Z\alpha)^{2}
\left[\log\left(\frac{2a}{\rho}\right)+\frac{(1-2\gamma)}{2}\right]\,,
\;\;\;\;\gamma=0.577\,.
\label{eq:2.210}
\eeq
For nuclei of finite radius $R_{A}$, Eq.~(\ref{eq:2.210})
holds
for $\rho\gsim R_{A}$, and $C(\rho\lsim
R_{A})=C(R_{A})$.
In section 2.4 we will give a more accurate
formula
for $C(\rho)$, which will be used in numerical
calculations.

The phase factor (\ref{eq:2.190}) is independent of 
$(\r_{i}+\r_{j})/2$. This allows one to calculate
the
path integral (\ref{eq:2.150}) analytically. This gives
the
following result
\cite{A2eBGZ}
\bea
S(\r_{2},\r_{2}',\xi_{2}|\r_{1},\r_{1}',\xi_{1})=
\left(\frac{\mu_{e}}{2\pi\Delta\xi}\right)^{2}
\exp\left\{\frac{i\mu_{e}}{2\Delta\xi}
[(\r_{1}-\r_{2})^{2}-(\r_{1}'-\r_{2}')^{2}]
\right.\nonumber\\
\left.
-\frac{1}{2}\int
d\xi
n(\xi)
\sigma(|\ta_{s}(\xi)|)\right\}\,,
\label{eq:2.220}
\eea
$$
\ta_{s}(\xi)=(\r_{1}-\r_{1}')\frac{(\xi_{2}-\xi)}{\Delta\xi}
+(\r_{2}-\r_{2}')\frac{(\xi-\xi_{1})}{\Delta\xi}\,,\;\;\;
\Delta\xi=\xi_{2}-\xi_{1}\,.
$$
In the case of the integral (\ref{eq:2.160}) we introduce
the
Jacobi
variables
$\bfa=(\mu_{e'}\r_{e'}+\mu_{\gamma}\r_{\gamma})/(\mu_{e'}+\mu_{\gamma})$
and $\r=\r_{e'}-\r_{\gamma}\,$. Then after analytical
path
integration over $\bfa(\xi)$ and $\r_{e}(\xi)$ in
Eq.
(\ref{eq:2.160}) we arrive
at
\bea
M(\r_{2},\r_{2}',\xi_{2}|\r_{1},\r_{1}',\xi_{1})=
\left(\frac{\mu_{e}}{2\pi\Delta\xi}\right)^{2}
\exp\left\{\frac{i\mu_{e}}{2\Delta\xi}
\left[(\r_{1}-\r_{2})^{2}-(\r_{1}'-\r_{2}')^{2}\right]
-\frac{i\Delta\xi}{L_{f}}\right\}\nonumber\\
\times
\int{\cal D}\r\exp\left\{i\int
d\xi\left[
\frac{\mu_{e'\gamma}\dot{\r}^{2}}{2}+i\frac{
n(\xi)
\sigma(|\ta_{m}(\xi)|)}{2}\right]\right\}
\,,
\label{eq:2.230}
\eea
$$
\ta_{m}(\xi)=(\r_{1}-\r_{1}')\frac{(\xi_{2}-\xi)}{\Delta\xi}
+(\r_{2}-\r_{2}')\frac{(\xi-\xi_{1})}{\Delta\xi}
+\frac{\r(\xi)\mu_{\gamma}}{(\mu_{e'}+\mu_{\gamma})}\,,
$$
where
$\mu_{e'\gamma}=\mu_{e'}\mu_{\gamma}/(\mu_{e'}+\mu_{\gamma})
=E_{e}x(1-x)$ is the reduced Schr\"odinger mass of the
$e'\gamma$
system.
It follows from Eqs. (\ref{eq:2.220}), (\ref{eq:2.230})
that
for the factors $S$ and $M$ there
hold
relations
\beq
\int d
\r_{2}
S(\r_{2},\r_{2},\xi_{2}|\r_{1},\r_{1}',\xi_{1})
=\delta(\r_{1}-\r_{1}')\,,
\label{eq:2.240}
\eeq
\beq
\int d
\r_{2}
M(\r_{2},\r_{2},\xi_{2}|\r_{1},\r_{1}',\xi_{1})
=\delta(\r_{1}-\r_{1}')
{\cal
K}(0,\xi_{2}|0,\xi_{1})\,,
\label{eq:2.250}
\eeq
where
\beq
{\cal
K}(\r_{2},\xi_{2}|\r_{1},\xi_{1})=
\int{\cal D}\r\exp\left\{i\int
d\xi\left[
\frac{\mu_{e'\gamma}\dot{\r}^{2}}{2}+
i\frac{n(\xi)\sigma(|\r
|x)}{2}
%v(\bfb,\xi)
\right]
\right\}
\label{eq:2.260}
\eeq
is the Green's function of
a
two-dimensional
Schr\"odinger equation with the
Hamiltonian
\beq
{\cal
H}=\frac{{\qb}^{2}}{2\mu_{e'\gamma}}+v(\r,\xi)\,,
\label{eq:2.270}
\eeq
\beq
v(\r,\xi)=-i\frac{n(\xi)\sigma(|\r
|x)}{2}\,.
\label{eq:2.280}
\eeq
The Hamiltonian (\ref{eq:2.270}) describes the
electron-photon
relative transverse
motion
in the $e'\gamma$ system. The form of the imaginary
potential
(\ref{eq:2.280})
reflects the fact that after integrating over
$\r_{2}$
in Eq. (\ref{eq:2.250}) the "positron"
trajectory
$\r_{e}(\xi)$ coincides with the trajectory of the
center-of-mass
of the $e'\gamma$
system.

Substituting (\ref{eq:2.220}), (\ref{eq:2.230})
into
(\ref{eq:2.120})
and integrating in
(\ref{eq:2.120})
over the transverse variables with the help of
(\ref{eq:2.240}),
(\ref{eq:2.250}), along with the normalization
condition
(\ref{eq:2.111}),
we finally obtain (we set $-z_{1}=z_{2}=\infty$, and recover
the
vacuum
term)
\beq
\frac{d P_{\gamma}}{d
x}=2\mbox{Re}
\int\limits_{-\infty}^{\infty} d
\xi_{1}
\int\limits_{\xi_{1}}^{\infty}d
\xi_{2}
\exp\left(-\frac{i\Delta\xi}{L_{f}}\right)
g(\xi_{1},\xi_{2},x)\left[{\cal
K}(0,\xi_{2}|0,\xi_{1})
-{\cal
K}_{v}(0,\xi_{2}|0,\xi_{1})
\right]\,.
\label{eq:2.290}
\eeq
Here ${\cal K}_{v}$ is the Green's function for the
Hamiltonian
(\ref{eq:2.270}) with
$v(\r,\xi)=0$.

Thus, we expressed the intensity of the photon
radiation
through the Green's function of the Schr\"odinger
equation
with the imaginary potential
(\ref{eq:2.280}).
Notice that the dipole cross section in Eq. (\ref{eq:2.280}) can
be
viewed
as an imaginary part of the forward scattering amplitude for
the
three-body $e^{+}e^{-}\gamma$
system.
In our analysis we neglected interaction of the
photon
with atomic electrons, which
becomes
important only for extremely soft photons
\cite{Migdal,TM}.
The inclusion of
this
interaction leads to appearance of a real part in the
potential
(\ref{eq:2.280}).

It is worth noting that an equation analogous
to
(\ref{eq:2.290}) holds also in more general case of
photon
emission on a random external potential if the photon
formation
length is much larger than the potential correlation
radius.
In this case the integral over $\xi$ in
Eq.
(\ref{eq:2.170}) defines a Gaussian random quantity. Using
the
formula $\langle \exp(iA)\rangle=\exp(-\langle A^{2}\rangle/2)$,
which is valid for a Gaussian random
quantity,
we find that the phase factor
(\ref{eq:2.170}) takes the form
\beq
\Phi(\{\r_{i}\},\{\r_{j}\})=
\exp\left[-i\int d
\xi
V\left(\r_{i}(\xi)-\r_{j}(\xi),\xi\right)\right]\,,
\label{eq:2.300}
\eeq
\beq
V(\r,\xi)=-2\pi
i\alpha\int\limits_{-\infty}^{\infty}d\xi'
\langle
\Delta
U(\r,\xi)U(\r,\xi')\rangle
\,,
\label{eq:2.310}
\eeq
where $\Delta
U(\r,\xi)=U(\r,\xi)-U(0,\xi)$.
The function (\ref{eq:2.310}) will play the role of the
imaginary
potential in the Hamiltonian
(\ref{eq:2.270}).
%
%-----------------------------------------------------------------
\subsection{ Bremsstrahlung in an infinite medium in
the oscillator
approximation}
%-----------------------------------------------------------------
To proceed with analytical evaluation of the
radiation
rate we take advantage of the slow
$\rho$-dependence
of $C(\rho x)$ at $\rho x\lsim 1/m_{e}$, which, as  will be seen
below, are important in
Eq.~(\ref{eq:2.290}).
Evidently, to a logarithmic accuracy we can
replace
(\ref{eq:2.280}) by the harmonic oscillator
potential
with the
frequency
\beq
\Omega=
\frac{(1-i)}{\sqrt{2}}
\left(\frac{n
C(\rho_{eff}x)x^{2}}{\mu_{e'\gamma}}\right)^{1/2}
=\frac{(1-i)}{\sqrt{2}}
\left(\frac{n
C(\rho_{eff}x)x}{E_{e}(1-x)}\right)^{1/2}
\,\,.
\label{eq:2.320}
\eeq
Here $\rho_{eff}$ is the typical value of $\rho$
for
trajectories dominating the radiation
rate.
Making use of the oscillator Green's
function
\beq
K_{osc}(\r_{2},z_{2}|\r_{1},z_{1})=
\frac{\mu\Omega}{2\pi i\sin{(\Omega\Delta
z)}}
\exp\left\{\frac{i\mu [(\r_{1}^{2}+\r_{2}^{2})\cos{(\Omega\Delta
z)}
-2\r_{1}\r_{2}]}
{2\sin{(\Omega\Delta
z)}}\right\}\,\,,
\label{eq:2.330}
\eeq
after
some
algebra one can obtain from Eq.~(\ref{eq:2.290})
the
intensity of bremsstrahlung per unit length in an infinite
medium
\beq
\frac{d P_{\gamma}}{dx
dL}=
n\left(\frac{C(\rho_{eff}x)}{C(1/m_{e})}\right)
\left[
\left(\frac{d\sigma}{dx}\right)^{BH}_{nf}
S_{nf}(\eta)
+\left(\frac{d\sigma}{dx}\right)^{BH}_{sf}S_{sf}(\eta)
\right]\,,
\label{eq:2.340}
\eeq
where $\eta=L_{f}|\Omega|$. In Eq. (\ref{eq:2.340}) we factored
out
the Bethe-Heitler cross
sections
conserving (nf) and changing (sf) the electron helicity, which
to
a logarithmic accuracy can be written
as
\beq
\left(\frac{d\sigma}{dx}\right)^{BH}_{nf}=
\frac{\alpha C(1/m_{e})}{3\pi
m_{e}^{2}}
\frac{[4-4x+2x^{2}]}{x}\,\,,
\label{eq:2.350}
\eeq
\beq
\left(\frac{d\sigma}{dx}\right)^{BH}_{sf}=
\frac{\alpha C(1/m_{e})}{3\pi
m_{e}^{2}}\,x\,\,.
\label{eq:2.351}
\eeq
The factors $S_{nf}$, $S_{sf}$ in Eq. (\ref{eq:2.340}) are given
by
\beq
S_{nf}(\eta)=\frac{3}{\eta\sqrt{2}}
\int\limits_{0}^{\infty}
dy
\left(\frac{1}{y^{2}}-
\frac{1}{{\rm sh}^{2}
y}\right)
\exp\left(-\frac{y}{\eta\sqrt{2}}\right)
\left[\cos\left(\frac{y}{\eta\sqrt{2}}\right)+
\sin\left(\frac{y}{\eta\sqrt{2}}\right)\right]\,,
\label{eq:2.360}
\eeq
\beq
S_{sf}(\eta)=\frac{6}{\eta^{2}}
\int\limits_{0}^{\infty}dy
\left(\frac{1}{y}-
\frac{1}{{\rm sh}
y}\right)
\exp\left(-\frac{y}{\eta\sqrt{2}}\right)
\sin\left(\frac{y}{\eta\sqrt{2}}\right)\,.
\label{eq:2.370}
\eeq
At small $\eta$ $S_{nf}(\eta)\simeq
1-16\eta^{4}/21$,
$S_{sf}(\eta)\simeq 1-31\eta^{4}/21$, and the Bethe-Heitler
regime
obtains. Up to the 
factor $C(\rho_{eff}x)/C(1/m_{e})$, which is slowly dependent on $\eta$,
the suppression
of
bremsstrahlung
at $\eta\gg 1$ is controlled by the asymptotic behavior of
the
suppression factors (\ref{eq:2.360}),
(\ref{eq:2.370}):
\beq
S_{nf}(\eta)\simeq
3/\eta\sqrt{2}\,,~~~~
S_{sf}(\eta)\simeq
3\pi/2\eta^{2}\,.
\label{eq:2.380}
\eeq
The value of $\rho_{eff}$
can
be estimated with the help
of
Eqs.~(\ref{eq:2.360}),
(\ref{eq:2.370}).
The variable of integration
in
(\ref{eq:2.360}),
(\ref{eq:2.370})
in terms of $\Delta \xi$ in Eq.~(\ref{eq:2.290}) equals $|\Delta
\xi
\Omega|$.
Therefore, for typical value of $\Delta \xi$ contributing
to
the integral (\ref{eq:2.290}), $\Delta\xi_{eff}$, we
have
$
\Delta\xi_{eff}\sim L_{f}^{'}={\mbox{
min}}(L_{f},1/|\Omega|)
$.
Note that $L_{f}^{'}$ plays the role of the
effective
medium-modified photon formation
length.
Having $\Delta\xi_{eff}$ we can estimate
$\rho_{eff}$
from the obvious Schr\"odinger diffusion
relation:
\beq
\rho_{eff}\sim
(2\Delta\xi_{eff}/\mu_{e'\gamma})^{1/2}\,.
\label{eq:2.390}
\eeq
In the low-density limit, when $\eta \rightarrow 0$, 
this relation yields $\rho_{eff}\sim 1/m_{e}x$, and the
right-hand
side of Eq. (\ref{eq:2.340}) goes over into the Bethe-Heitler
cross
section times the target
density.
In the soft
photon
limit ($x\rightarrow 0$) at fixed $n$, $\eta$ becomes much greater
than
unity. In this regime of strong LPM suppression, using 
the asymptotic formula for $S_{nf}$, one can obtain from
Eqs.
(\ref{eq:2.340}),
(\ref{eq:2.350})
\beq
\frac{d P_{\gamma}}{dx
dL}\approx
2\alpha^{2}Z\,\sqrt{\frac{2 n\log(2a/\rho_{eff}x)}{\pi
E_{e}x}}\,,
\label{eq:2.400}
\eeq
with
$
\rho_{eff}\sim
\left[
\pi (Z\alpha)^{2} n E_{e}x^{3}\log(2/\alpha
Z^{1/3})\right]^{-1/4}\,.
$
This result agrees with Migdal's prediction \cite{Migdal}
obtained
within the Fokker-Planck approximation in the momentum
representation.
Our suppression factors (\ref{eq:2.360}), (\ref{eq:2.370})
also
agree with those obtained in Ref. \cite{Migdal}. Equivalence
of
the the oscillator approximation
in
coordinate representation
to
the Fokker-Planck one in momentum representation is
not
surprising. Making use
of
Eq. (\ref{eq:2.220})
one can easily show that $\sigma(\rho)\propto \rho^{2}$
leads
to the Gaussian diffusion in the momentum space. That is, we
have
a diffusion described by the Fokker-Planck
equation.

The oscillator approximation simplifies greatly
evaluation
of the radiation rate. It allows one to obtain simple
formulas
for suppression factors for a finite-size target as
well.
The corresponding analysis within Migdal's approach
was
performed in
Ref. \cite{Tern}.
Unfortunately, the oscillator
approximation is accurate only for strong LPM suppression, when one
can
neglect the variation of the factor
$C(\rho)$.
This effect must be taken into account to evaluate accurately 
the radiation rate in an infinite
medium
in the regime of small LPM suppression and
for
finite-size
targets.
In the next section we represent
Eq. (\ref{eq:2.290})
in a different form which is more convenient for numerical
calculations
with a rigorous treatment of the Coulomb
effects.

%---------------------------------------------------------------------
\subsection{ Glauber form of the radiation rate}
%---------------------------------------------------------------------
In this section we demonstrate that Eq. (\ref{eq:2.290}) can
be
rewritten in a form analogous to the Glauber amplitude for
elastic
hadron-nucleus
scattering.
Let us expand the Green's function ${\cal K}$ in
Eq. (\ref{eq:2.290})
in a series in the potential
$v$
$$
{\cal
K}(\r_{2},z_{2}|\r_{1},z_{1})=
{\cal
K}_{v}(\r_{2},z_{2}|\r_{1},z_{1})
\,\,\,\,\,\,\,\,\,\,\,\,\,\,\,\,\,\,\,\,\,\,\,\,\,\,\,\,\,\,\,\,\,\,\,
\,\,\,
$$
$$
\,\,\,\,\,\,\,\,\,\,\,\,\,\,\,\,\,\,\,\,\,\,\,\,
+\int\limits_{z_{1}}^{z_{2}}dz\int
d\r
{\cal
K}_{v}(\r_{2},z_{2}|\r,z)(-iv(\r,z))
{\cal
K}_{v}(\r,z|\r_{1},z_{1})\,+\cdots\,.
$$
Then after a simple algebra one can represent (\ref{eq:2.290})
in
the
form
\beq
\frac{d P_{\gamma}}{d
x}=
\frac{d P_{\gamma}^{BH}}{d x}+\frac{d P_{\gamma}^{abs}}{d
x}\,,
\label{eq:2.410}
\eeq
where
\bea
\frac{d P_{\gamma}^{BH}}{d
x}=-T\cdot\mbox{Re}
\int
d\r
\int\limits_{-\infty}^{0} d\xi_{1}\int\limits_{0}^{\infty}
d\xi_{2}
g(\xi_{1},\xi_{2},x)
{\cal
K}_{v}(0,\xi_{2}|\r,0)\nonumber\\
\times\sigma(\rho x){\cal
K}_{v}(\r,0|0,\xi_{1})
\exp\left[-\frac{i(\xi_{2}-\xi_{1})}{L_{f}}\right]\,,
\label{eq:2.420}
\eea
\bea
\frac{d P_{\gamma}^{abs}}{d
x}=\frac{1}{2}\mbox{Re}
\int\limits_{0}^{L}
dz_{1}n(z_{1})
\int\limits_{z_{1}}^{L}
dz_{2}n(z_{2})
\int
d\r_{1}d\r_{2}
\int\limits_{-\infty}^{z_{1}}
d\xi_{1}\int\limits_{z_{2}}^{\infty}
d\xi_{2}
g(\xi_{1},\xi_{2},x)
{\cal
K}_{v}(0,\xi_{2}|\r_{2},z_{2})\nonumber\\
\times\sigma(\rho_{2} x){\cal
K}(\r_{2},z_{2}|\r_{1},z_{1})
\sigma(\rho_{1}x)
{\cal
K}_{v}(\r_{1},z_{1}|0,\xi_{1})
\exp\left[-\frac{i(\xi_{2}-\xi_{1})}{L_{f}}\right]\,.
\label{eq:2.430}
\eea
Here
$
T=\int_{0}^{L} dz
n(z)
$
is the optical thickness of the target (we assume that $n(z)=0$ at
$z<0$
and
$z>L$).
The integrals over $\xi_{1,2}$
in
(\ref{eq:2.420}),
(\ref{eq:2.430})
of the products of the vacuum Green's functions and
exponential
phase factors can be expressed through the
light-cone
wave
function
$
\Psi(x,\r,\lambda_{e},\lambda_{e'},\lambda_{\gamma})
$
for the transition $e\rightarrow e'\gamma$. At
$\lambda_{e'}=\lambda_{e}$
it
is
\bea
\Psi(x,\r,\lambda_{e},\lambda_{e'},\lambda_{\gamma})=
\frac{-i}{2\mu_{e'\gamma}}\sqrt{\frac{\alpha}{2x}}
\left[\lambda_{\gamma}(2-x)+2\lambda_{e}x\right]
\left(\frac{\partial}{\partial
\rho_{x}}-
i\lambda_{\gamma}\frac{\partial}{\partial
\rho_{y}}\right)
\int\limits_{-\infty}^{0}d\xi {\cal
K}_{v}(\r,0|0,\xi)
\nonumber \\
\times
\exp\left(\frac{i\xi}{L_{f}}\right)=
\frac{1}{2\pi}\sqrt{\frac{\alpha
x}{2}}
\left[\lambda_{\gamma}(2-x)+2\lambda_{e}x\right]
\exp(-i\lambda_{\gamma}\varphi)m_{e}K_{1}(\rho
m_{e}x)\,,
\label{eq:2.440}
\eea
for $\lambda_{e'}=-\lambda_{e}$ the only
nonzero
component is the one with
$\lambda_{\gamma}=2\lambda_{e}$
\bea
\Psi(x,\r,\lambda_{e},-\lambda_{e},2\lambda_{e})=
\frac{\sqrt{2\alpha
x^{3}}}{2\mu_{e'\gamma}}
\int\limits_{-\infty}^{0} d\xi{\cal
K}_{v}(\r,0|0,\xi)
\exp\left(\frac{i
\xi}{L_{f}}\right)
=\frac{-i}{2\pi}\sqrt{2\alpha x^{3}}m_{e}K_{0}(\rho
m_{e}x)\,.
\label{eq:2.450}
\eea
Here $K_{0}$ and $K_{1}$ are the Bessel
functions.
Eqs. (\ref{eq:2.440}), (\ref{eq:2.450}) can be obtained
by
calculating the matrix
element
for transition $e\rightarrow
e'\gamma$
in time-ordered PT using the representation
(\ref{eq:2.30})
for electron and photon wave
functions.

Making use of Eqs.~(\ref{eq:2.440}),~(\ref{eq:2.450}) one can
rewrite
(\ref{eq:2.420}),~(\ref{eq:2.430}) in the
form
\beq
\frac{d P_{\gamma}^{BH}}{d
x}=\frac{T}{2}
\sum\limits_{\{\lambda_{i}\}} \int
d\r\,
|\Psi(x,\r,\{\lambda_{i}\})|^{2}
\sigma(\rho
x)\,\,,
\label{eq:2.460}
\eeq
\bea
\frac{d P_{\gamma}^{abs}}{d
x}=-\frac{1}{4}\mbox{Re}
\sum\limits_{\{\lambda_{i}\}}
\int\limits_{0}^{L}
dz_{1}n(z_{1})
\int\limits_{z_{1}}^{L}
dz_{2}n(z_{2})
\int
d\r\,
\Psi^{*}(x,\r,\{\lambda_{i}\})\nonumber\\
\times\sigma(\rho
x)
\Phi(x,\r,\{\lambda_{i}\},z_{1},z_{2})
\exp\left[-\frac{i(z_{2}-z_{1})}{L_{f}}\right]\,,
\label{eq:2.470}
\eea
where
\beq
\Phi(x,\r,\{\lambda_{i}\},z_{1},z_{2})=
\int d\r'{\cal
K}(\r,z_{2}|\r',z_{1})
\Psi(x,\r',\{\lambda_{i}\})\,\sigma(\rho'
x)
\label{eq:2.480}
\eeq
is the solution of the Schr\"odinger equation with the
boundary
condition
$
\Phi(x,\r,\{\lambda_{i}\},z_{1},z_{1})=
\Psi(x,\r,\{\lambda_{i}\})\sigma(\rho
x)\,.
$

In Ref. \cite{NPZ} it was shown that the
$p_{\perp}$-integrated
cross
section
for a radiation process $a\rightarrow b
c$
can be written
as
\beq
\frac{d\sigma ({a\rightarrow
cb})}{dx}=
\int d\r
\,W_{a}^{bc}(x,\r)
\sigma_{\bar{a}bc}(\rho)\,,
\label{eq:2.490}
\eeq
where $W_{a}^{bc}(x,\r)=|\Psi_{a}(x,\r)|^{2}$ is the
light-cone
probability
distribution
for transition $a\rightarrow bc$, $\sigma_{\bar{a}bc}$ is the
total
cross section of interaction with the target of $\bar{a}bc$
system.
For the transition $e\rightarrow e'\gamma$ the
corresponding
three-body cross section equals $\sigma(\rho
x)$. Consequently,
the first term in (\ref{eq:2.410}) equals the
Bethe-Heitler
cross section times the target optical thickness, {\it i.e.}
it
corresponds to the impulse
approximation,
while the second term describes LPM
suppression.
Thus, we have demonstrated that LPM suppression is
equivalent
to absorption for $e^{+}e^{-}\gamma$
system.

It is worth noting that at $L_{f}\gg L$ the 
radiation
rate for a composite target can also be represented in a form similar to
that of Eq.
(\ref{eq:2.490}).
Indeed, in this limit the transverse variable $\r$ is
approximately
frozen, and the Green's function can be
written
in the eikonal
form
\beq
{\cal
K}(\r_{2},z_{2}|\r_{1},z_{1})\approx
\delta(\r_{2}-\r_{1})\exp
\left[-\frac{\sigma(\rho_{1}x)}{2}
\int\limits_{z_{1}}^{z_{2}}
dz\,n(z)
\right]\,.
\label{eq:2.500}
\eeq
Using (\ref{eq:2.500}) we obtain in the frozen-size approximation
from
Eqs.~(\ref{eq:2.410}),~(\ref{eq:2.460}),~(\ref{eq:2.470})
\beq
\frac{dP_{\gamma}^{fr}}{dx}=
2\int d\r
\,W_{e}^{e\gamma}(x,\r)
\left\{1-\exp\left[-\frac{T\sigma(\rho
x)}{2}\right]\right\}\,.
\label{eq:2.510}
\eeq
Eq. (\ref{eq:2.510}) is analogous to the formula obtained in Ref. \cite{NPZ}
for the
cross
section of heavy quark production in hadron-nucleus collision.
Within classical electrodynamics the
LPM
effect at $L_{f}\gg L$ was previously discussed in
Ref.
\cite{ShF}.

Representation (\ref{eq:2.410}) has the virtue of
bypassing
calculation
of the singular transverse Green's function. This renders
it
convenient for numerical calculations of the radiation rate
for
finite-size
targets.

%-------------------------------------------------------------------
\subsection{ Numerical results and comparison with the SLAC
experiment}
%-------------------------------------------------------------------
For numerical calculations we need a more
accurate
parametrization of the dipole cross section, which
takes
into account the inelastic processes and the Coulomb
correction.
We write the dipole cross section in the
form
$
\sigma(\rho)=\rho^{2}C(\rho)\,,
$
where
\beq
C(\rho)=Z^{2}C_{el}(\rho)+Z
C_{in}(\rho)\,.
\label{eq:2.521}
\eeq
Here the terms $\propto Z^{2}$ and $\propto Z$ correspond
to
elastic and inelastic intermediate states in
interaction
of $e^{+}e^{-}$ pair with an
atom.
Due to the steep decrease of the light-cone wave
function
$\Psi(x,\r,\{\lambda_{i}\})$ at $|\r|\gsim
1/m_{e}x$
the
dominating values of $\rho$ in (\ref{eq:2.460}) are $\sim
1/m_{e}$.
For (\ref{eq:2.470}) they are even smaller due to the
absorption
effects.
For this reason the probability of photon emission is only
sensitive
to the behavior of $\sigma(\rho)$ at $\rho\lsim 1/m_{e}\ll
r_{B}$.
In this region
both
the $C_{el}$ and $C_{in}$ can have only weak logarithmic
dependence
on
$\rho$.
This allows one to parametrize them in the
form
\bea
C_{i}(\rho)=8\pi\left(\frac{\alpha
a_{i}}{\rho}\right)^{2}
\left[1-\frac{\rho}{a_{i}}K_{1}\left(\frac{\rho}{a_{i}}\right)\right]
\approx
4\pi\alpha^{2}
\left[\log\left(\frac{2a_{i}}{\rho}\right)+\frac{(1-2\gamma)}{2}
\right]\,.
\label{eq:2.530}
\eea
For elastic
component
$C_{el}(\rho\lsim
R_{A})=C_{el}(R_{A})$.
We adjusted $a_{el}$ and
$a_{in}$
to reproduce the terms $\propto Z^{2}$ and $\propto Z$,
respectively,
in the Bethe-Heitler cross
section
\beq
\frac{d\sigma^{BH}}{dx}=
\frac{4\alpha^{3}}{3
m_{e}^{2}x}
\left\{(4-4x+3x^{2})[Z^{2}(F_{el}-f(Z\alpha))+Z
F_{in}]
+(1-x)\frac{(Z^{2}+Z)}{3}\right\}\,,~~~~~
\label{eq:2.540}
\eeq
$$
F_{el}\approx
\log(184/Z^{1/3})\,,~~~~~~~~~~
F_{in}\approx
\log(1194/Z^{2/3})\,,
$$
$$
f(y)=y^{2}\sum\limits_{n=1}^{\infty}\frac{1}{n(n^{2}+y^{2})}
$$
evaluated
in the standard approach with realistic atomic formfactors
\cite{Tsai}.
This procedure
gives
$a_{el}=0.81\, r_{B}Z^{-1/3}\exp(-f(Z\alpha))$
and
$a_{in}=5.2\,
r_{B}Z^{-2/3}$.

In Fig. 1 we compare the results of
calculations
(solid curve)
of
the bremsstrahlung rate with the one measured in
\cite{SL1}
for a gold target with $L=0.7\%X_{0}\approx 0.023$
mm
($X_{0}$ is the radiation length) and 25 GeV electron
beam.
We also show the prediction of
frozen-size
approximation (\ref{eq:2.510}) (dashed
curve),
the radiation rate obtained for the infinite
medium
(long-dashed curve), and the Bethe-Heitler
spectrum
(dot-dashed curve). We have found that the normalization
of
the experimental spectrum disagrees a little with our
theoretical
prediction. The theoretical curves in Fig.~1
were
multiplied by the factor 1.03. This
renormalization
brings the calculated spectrum in very good agreement with the
data
of
Ref. \cite{SL1}.
\footnote{
Recently we have analyzed the SLAC data \cite{SL1} including
the multiphoton effects ({\sl hep-ph}/9805271). The results of this
analysis are in very good
agreement with the 
experimental data
for all the targets used in \cite{SL1}. For the 0.7\%$X_{0}$ gold target
the effect of multiphoton emission turns out to be small.
It increases the normalization
constant by $\sim 3$\%.}

For 25 GeV
electrons
$L_{f}\approx 0.94\cdot (1 \mbox{MeV}/k(\mbox{MeV}))$ mm in
the
region
of
$k$ shown in Fig.~1. One can conclude from this figure that the
radiation
density calculated using
Eqs.
(\ref{eq:2.410}),~(\ref{eq:2.460}),~(\ref{eq:2.470})
is close to the prediction of the
frozen-size
approximation (\ref{eq:2.510}) for the photons with $L_{f}\gsim 2
L$,
while for the photons with $L_{f}\lsim L$ it is close to
the
spectrum for the infinite
medium.
To illustrate the role of the finite target thickness better we
present
in Fig.~2 the LPM suppression factor defined
as
$S=\frac{dP_{\gamma}/dx}{dP_{\gamma}^{BH}/dx}$
as a function of the ratio $h=L/L_{f}$ for several values
of
the photon momentum. The calculations were performed for a gold target
and
25 GeV electron
beam.
Fig.~2 demonstrates that the edge effects come into
play
at $L\lsim L_{f}$. One can also see from Fig.~2 that
for
low photon momenta the edge effects vanish
steeper.
This fact is a consequence of a stronger suppression of the
coherence
length in radiation of soft
photons.

Fig.~2 shows that the suppression factor
has
a minimum at $L\approx L_{f}$ for 100 and 400 MeV
photons.
This minimum reflects the two-edge
interference
for a plate
target.
One can expect a more pronounced interference
effects
for structured
targets.
To illustrate the role of the interference effects in
Fig.~3
we show our results for the LPM suppression factor for a
two
segment gold
target.
We have performed calculations for the same plate
thicknesses
and the gaps between two plates as in the recent paper
by
Blankenbecler \cite{Blan}.  The analysis \cite{Blan}
was
performed within the model proposed in Ref. \cite{BlanD},
in
which the medium was modelled by the
potential
$
U(\r,z)=-\r\cdot \mbox{\bf
E}_{\perp}(z)
$
,
where $\mbox{{\bf E}}_{\perp}$ is a random transverse electric
field.
Qualitatively our results are similar to those of
Ref. \cite{Blan}.
However, for our realistic electron-atom interaction
the
maxima and minima in the spectra are less pronounced than
for
the model medium used in
Ref. \cite{Blan}.
For a homogeneous target our spectrum differs from
obtained
by Blankenbecler by $\sim
10-20$\%.

The disagreement of our results with those
of
Blankenbecler
is a consequence of impossibility to simulate the
Coulomb
effects in the approach of
Refs. \cite{BlanD,Blan}.
Indeed, using Eq. (\ref{eq:2.310})
one
can show that the model potential of
Refs. \cite{BlanD,Blan}
corresponds in our approach to the following choice of the
dipole
cross
section
$$
\sigma(\rho)=\frac{2\pi
\alpha\rho^{2}}{n}\int\limits_{-\infty}^{\infty}
d z \,\langle \mbox{\bf
E}_{\perp}(0)\cdot
\mbox{\bf
E}_{\perp}(z)
\rangle\,.
$$
Thus we see that in the approach of Ref. \cite{Blan} the
Coulomb
effects, leading to the important logarithmic $\rho$-dependence of
the
factor $C(\rho)$ (\ref{eq:2.521}), are
missed.
We conclude that the model of Refs. \cite{BlanD,Blan} is too
crude
for a quantitative simulation of the LPM effect in a real
medium.

%-----------------------------------------------------------
\subsection{ Probability of $e^{+}e^{-}$ pair production}
%-----------------------------------------------------------

The probability of pair production by a high energy
photon
can be written in the form similar to
Eq. (\ref{eq:2.290}).
In this
case
the two-dimensional Hamiltonian
reads
\beq
{\cal H}=\frac{{\bf
q}^{2}}{2\mu_{e\bar{e}}}+v(\r,\xi)\,,
\label{eq:2.541}
\eeq
\beq
v(\r,\xi)=-i\frac{n(\xi)\sigma(|\r
|)}{2}\,,
\label{eq:2.550}
\eeq
where $\mu_{e\bar{e}}=E_{\gamma}x(1-x)$, $x$ is
the
electron fractional light-cone
momentum.
The formation length for pair production
is
$L_{f}={2E_{\gamma}x(1-x)}/{m_{e}^{2}}\,$,
and
the vertex operator is given
by
\beq
g(\xi_{1},\xi_{2},x)=
\frac{\alpha[x^{2}+(1-x)^{2}]}{2}\,
\vb(\xi_{2})\cdot\vb(\xi_{1})+
\frac{\alpha
m_{e}^{2}}{2\mu_{e\bar{e}}^{2}}\,\,,
\label{eq:2.560}
\eeq
where
$$
\vb(\xi_{i})={\vb}_{e}(\xi_{i})
-{\vb}_{\bar{e}}(\xi_{i})\,,
$$
$\vb_{e}$ and $\vb_{\bar{e}}$ are the electron and
positron
transverse velocity
operators.

The light-cone wave function for
transition
$\gamma\rightarrow e^{+}e^{-}$ entering the representation
analogous
to Eq. (\ref{eq:2.410}) is as
follows:
\bea
\Psi(x,\r,\lambda_{\gamma},\lambda_{e},\lambda_{\bar{e}})=
\frac{1}{2\pi}\sqrt{\frac{\alpha
}{2}}
\left[\lambda_{\gamma}(2x-1)+2\lambda_{e}\right]
\exp(i\lambda_{\gamma}\varphi)m_{e} K_{1}(\rho
m_{e})\,,
\label{eq:2.570}
\eea
for
$\lambda_{e}=-\lambda_{\bar{e}}$,
and the only nonzero
component
for
$\lambda_{e}=\lambda_{\bar{e}}$
(in this case $\lambda_{\gamma}=2\lambda_{e}$)
is
\beq
\Psi(x,\r,\lambda_{\gamma},\lambda_{e},\lambda_{\bar{e}})=
\frac{i}{2\pi}\sqrt{2\alpha }m_{e}K_{0}(\rho
m_{e})\,.
\label{eq:2.580}
\eeq

%---------------------------------------------------------------------
\section{The LPM effect in QCD}
%---------------------------------------------------------------------
\subsection{  General expression for the probability of gluon
emission}
%---------------------------------------------------------------------
Let us now consider the LPM effect for the induced gluon
radiation
from a fast
quark.
We discuss both cold nuclear matter and
QGP.
For QGP we use the GW model \cite{GW1} treating
QGP
as a system of static scattering centers described
by
the Debye screened
potential.
For the Debye color
screening
mass we use perturbative
formula
$\mu_{D}=(1+n_{F}/6)^{1/2}
g_{s}\,T$
\cite{Shuryak1}, where $g_{s}=\sqrt{4\pi \alpha_{s}}$ is the
QCD
coupling constant, $T$ is the temperature of
QGP.
Nucleons making up the cold nuclear matter are also treated
as
static scattering
centers.
Interaction of the fast quark and emitted gluon
with
each center will be described including one- and
two-gluon
exchanges. It should be noted that inclusion of the
two-gluon
exchange is absolutely necessary to ensure unitarity.

The derivation of the gluon radiation rate follows closely
the
analysis of bremsstrahlung in
QED.
Similarly to Eq. (\ref{eq:2.20}), the probability of gluon
emission,
$P_{g}$,
is connected with the medium modification of the radiative
correction
to the probability to detect in the final state one quark, $\delta
P_{q}$,
\beq
P_{g}=-(\delta P_{q}-\delta
P_{q}^{vac})\,.
\label{eq:3.10}
\eeq
Owing to the fact that $-T^{*}_{q}=T_{\bar{q}}$
(here
$T_{q,\bar{q}}$ are the color generators for a quark and
an antiquark)
the complex conjugated quark propagator is equivalent
to
the antiquark propagator. After summing over the final states of
the
target with the help of the closure
relation
$$
|\Psi_{t}^{f}\rangle\langle
\Psi_{t}^{f}|=1\,,
$$
where
$\Psi_{t}^{f}$ is the wave function of the target after
interaction
with a fast quark, the $\delta P_{q}$ will involve only the
diagonal
matrix elements for the medium constituents. This means
that
only
the diagrams involving color singlet (Pomeron)
$t$-channel
exchanges between the $q\bar{q}$, $q\bar{q}g$ states and the
medium
constituents contribute to $\delta
P_{q}$.
Consequently, in just the same way as in 
QED, we can obtain the expression for $\delta P_{q}$ in a
medium
introducing the corresponding absorption
factor
in the vacuum path integral formula for $\delta
P_{q}$.
This allows one to obtain the formulas analogous to
Eqs.
(\ref{eq:2.120}), (\ref{eq:2.150}) and
(\ref{eq:2.160}).
In the analogue of
Eq.
(\ref{eq:2.150})
the
absorption
factor contains the dipole cross section $\sigma_{2}$ of
interaction
of $q\bar{q}$ pair with a medium constituent. In the
absorption
factor
for
the QCD analogue of Eq. (\ref{eq:2.160}) the corresponding
cross
section
is the three-body cross section $\sigma_{3}$ for
$q\bar{q}g$
intermediate state. Namely this cross section enters the
final
formula for the radiation rate. For a quark incident
on
a target, it has a form that is analogous to equation (\ref{eq:2.290})
(we use notation similar to that in the case of
QED)
\beq
\frac{d P_{g}}{d
x}=2\mbox{Re}\!
\int\limits_{-\infty}^{\infty}\! d
\xi_{1}\!
\int\limits_{\xi_{1}}^{\infty}d
\xi_{2}
\exp\left(-\frac{i\Delta
\xi}{L_{f}}\right)
g(\xi_{1},\xi_{2},x)\left[{\cal
K}(0,\xi_{2}|0,\xi_{1})
-{\cal
K}_{v}(0,\xi_{2}|0,\xi_{1})\right]\,.
\label{eq:3.20}
\eeq
Here the generalization of the QED vertex operator
(\ref{eq:2.100})
to QCD
reads
\beq
g(\xi_{1},\xi_{2},x)=
\frac{\alpha_{s}[4-4x+2x^{2}]}{3x}\,
\vb(\xi_{2})\cdot\vb(\xi_{1})
+ \frac{2\alpha_{s} m_{q}^{2}x}{3\mu_{q'}^{2}}
\,
\,.
\label{eq:3.30}
\eeq
where
$$
\vb(\xi_{i})={\vb}_{g}(\xi_{i})
-{\vb}_{q'}(\xi_{i})\,,
$$
$\vb_{g}$ and $\vb_{q'}$ are the gluon and
quark
transverse velocity
operators.
The Hamiltonian for the Green's
function
${\cal K}$ is given
by
\beq
{\cal
H}=\frac{{\qb}^{2}}{2\mu_{q'g}}+v(\r,z)\,,
\label{eq:3.40}
\eeq
\beq
v(\r,z)=-i\frac{n(z)\sigma_{3}(\rho,x)}{2}\,.
\label{eq:3.50}
\eeq
The Schr\"odinger masses are defined
similarly
to the case of photon
radiation.
The gluon formation length
is
\beq
L_{f}=\frac{2E_{q}x(1-x)}{[m_{q}^{2}x^{2}+m_{g}^{2}(1-x)]}\,\,,
\label{eq:3.60}
\eeq
$m_{q}$ is the quark mass and $m_{g}$ is the mass of
radiated
gluon. The latter plays the
role
of an infrared cutoff removing contribution of the long-wave
gluon
excitations which cannot propagate in the real nonperturbative
QCD
vacuum.
In the case of QGP
summation over triplet (quark) and octet
(gluon)
color states is implied on the right-hand side of Eq. (\ref{eq:3.50}).
For a quark produced inside a medium through a hard mechanism
the
integration over $\xi_{1}$ in Eq. (\ref{eq:3.20}) starts from
the
production
point.
Note that for gluon
emission
$L_{f}\rightarrow 0$ for soft ($x\rightarrow
0$)
and hard ($x\rightarrow 1$) gluons. As a result, in both
these
limiting cases the Bethe-Heitler regime must
obtain.

In general case the three-body cross section for $q\bar{q}g$
state
depends on the two transverse vectors: $\r_{qg}$
and
$\r_{\bar{q}g}$,
here $\r_{ij}=\r_{i}-\r_{j}$. In terms of the dipole
cross
section it is given by
\cite{NZJETP78}
\beq
\sigma_{3}(\r_{qg},\r_{\bar{q}g})=
\frac{9}{8}[\sigma_{2}(|\r_{qg}|)+\sigma_{2}(|\r_{\bar{q}g}|)]
-\frac{1}{8}\sigma_{2}(|\r_{q\bar{q}}|)\,.
\label{eq:3.70}
\eeq
However, in the case of interest
the
antiquark in the $q\bar{q}g$ system is located at the
center-of-mass
of the $qg$ system,
and
\beq
\r_{\bar{q}g}=(1-x)\r_{qg}\,,~~~~~\r_{q\bar{q}}=-x\r_{qg}\,.
\label{eq:3.80}
\eeq
For this reason the three-body cross section entering
the
imaginary potential (\ref{eq:3.50}) can be written
as
\beq
\sigma_{3}(\rho,x)=\frac{9}{8}[\sigma_{2}(\rho)
+\sigma_{2}((1-x)\rho))]-\frac{1}{8}\sigma_{2}(x\rho)\,,
\label{eq:3.90}
\eeq
where $\rho=|\r_{qg}|$ . Eqs. (\ref{eq:3.80}),
(\ref{eq:3.90})
demonstrate that at $x\rightarrow 0$
the
color singlet $q\bar{q}g$ system interacts with medium
constituents
as octet-octet state, and as triplet-triplet state at $x\rightarrow
1$. This
is a direct consequence of the $x$-dependence of the
transverse
separations defined by Eq. (\ref{eq:3.80}). Notice
that
this makes evident that in the soft gluon limit one cannot
neglect
the transverse size of the $qg$ system as was done
in
Refs. \cite{B1,Levin,B3}.

The dipole cross section can be written
as
\beq
\sigma_{2}(\rho)=C_{2}(\rho)\rho^{2}\,,
\label{eq:3.100}
\eeq
where $C_{2}(\rho)$ has a smooth (logarithmic) dependence
on
$\rho$ at small $\rho$. For nucleon $C_{2}(\rho)$ in the
small-$\rho$
limit can be expressed through the gluon distribution
\cite{NZxg}
\beq
C_{2}(\rho)\approx
\frac{\pi^{2}\alpha_{s}(\rho)}{3}x_{B}g(x_{B},Q^{2}\approx
\frac{{\cal A}}{\rho^{2}})\,,~~~~~~~~{\cal A}\approx
10\,,
\label{eq:3.110}
\eeq
where $x_{B}\sim (2xE_{q}m_{p}\rho^{2})^{-1}$. For energies that are of
interest
from the practical viewpoint, the gluon density in
Eq.
(\ref{eq:3.110}) can be estimated in the Born approximation,
which
corresponds
to calculation of $\sigma_{2}$ in the double gluon model of
the
Pomeron
\cite{DGM}.

It is appropriate here to comment on gluon emission by a
fast
gluon. In this case $q\bar{q}g$ state will be replaced
by
$ggg$ state. The $ggg$ system can be in symmetric
and
antisymmetric color
states.
As a result, the cross section for $q\bar{q}g$ state in
the
potential (\ref{eq:3.50}) will be replaced by the
diffraction
operator describing transitions between these two color
states.
However, in soft gluon limit
the
transition to symmetric color state can be neglected
and
we obtain the same Schr\"odinger equation as for gluon
emission
by a quark. The corresponding vertex operator is given
by
Eq. (\ref{eq:3.30}) times the color factor
9/4.

In this study  we
will
use formula (\ref{eq:3.20}) to evaluate the
quark
energy
loss
\beq
\Delta E_{q}=E_{q}\int\limits_{0}^{1} dx
x\frac{dP_{g}}{dx}\,.
\label{eq:3.120}
\eeq
We will consider
homogeneous
nuclear matter and QGP. Of course, due to dependence of
the
probability
of gluon emission on gluon and quark masses, our
theoretical
predictions are of approximate, estimating
nature.
Bearing this in mind, we will neglect
the
spin-flip transitions, which give a small
contribution
to the energy loss. Note that, in any case for a quark produced
through
a hard mechanism inclusion of spin effects in radiation
without
those at production vertex does not make
sense.

%------------------------------------------------------------------
\subsection{ Gluon emission in an
infinite medium in the oscillator
approximation}
%------------------------------------------------------------------
Using Eq. (\ref{eq:2.490}) for the transition $q\rightarrow qg$
one
can show that the Bethe-Heitler cross section
is
dominated
by
the contribution
from
$\rho\lsim 1/m_{g}$. In the case of gluon emission in a
medium,
the typical values of the transverse separations in the
$q\bar{q}g$
system are still smaller due to absorption of the
configurations
with large transverse size
.
The smooth $\rho$-dependence of $C_{2}(\rho)$ at
$\rho\lsim
1/m_{g}$ allows one to evaluate
the
induced gluon radiation to a logarithmic accuracy
replacing
$C_{2}(\rho)$
by
$ C_{2}(\rho_{eff})$, where $\rho_{eff}$ is the typical
size
of the $q\bar{q}g$ system dominating the radiation
rate
(\ref{eq:3.20}).
Then
$\sigma_{3}(\rho,x)\approx C_{3}(x)\rho^{2}$,
where
\beq
C_{3}(x)=\frac{1}{8}\left\{9[1+(1-x)^{2}]-x^{2}\right\}
C_{2}(\rho_{eff})\,,
\label{eq:3.121}
\eeq
and the Hamiltonian
(\ref{eq:3.40})
takes the oscillator form with
the
frequency
$$
\Omega=\frac{(1-i)}{\sqrt{2}}
\left(\frac{nC_{3}(x)}{\mu_{q'g}}\right)^{1/2}=
\frac{(1-i)}{\sqrt{2}}
\left(\frac{nC_{3}(x)}{E_{q}x(1-x)}\right)^{1/2}\,.
$$
Note that the large value of factor ${\cal A}$ in
Eq. (\ref{eq:3.110}) is important 
from the viewpoint of applicability of
the
oscillator approximation.  This
allows one to use this approximation for a qualitative analysis
of
the
induced gluon radiation even for a weak LPM
effect
when $\rho_{eff}\sim
1/m_{g}$.

Using the oscillator Green's function
(\ref{eq:2.330}), we can obtain for the radiation rate per
unit
length
\beq
\frac{d
P_{g}}{dxdL}=
n\frac{d\sigma^{BH}_{}}{dx}S(\eta)\,,
\label{eq:3.130}
\eeq
where the suppression factor is defined by
Eq. (\ref{eq:2.360}),
and the Bethe-Heitler cross section is given
by
\beq
\frac{d\sigma^{BH}_{}}{dx}=
\frac{4\alpha_{s}C_{3}(x)(4-4x+2x^{2})}{9\pi
x[m_{q}^{2}x^{2}+m_{g}^{2}(1-x)]}\,.
\label{eq:3.140}
\eeq
The dimensionless parameter $\eta$ in (\ref{eq:3.130})
reads
\beq
\eta=L_{f}|\Omega|=
\frac{[4 n C_{3}(x)
E_{q}x(1-x)]^{1/2}}{m_{q}^{2}x^{2}+m_{g}^{2}(1-x)}
\,\,.
\label{eq:3.160}
\eeq

Note that the Bethe-Heitler cross section has
the
infrared $1/m_{g,q}^{2}$
divergence.
However, it is interesting that,
in
the
limit
of strong LPM suppression $\eta>>1$, multiple scattering
eliminates
this
divergence.
Using the asymptotic formula (\ref{eq:2.380}) for $S_{nf}(\eta)$
at
$\eta >>1$, we
can
obtain from Eqs. (\ref{eq:3.130}), (\ref{eq:3.140}) in this
regime
\beq
\frac{d P_{g}}{dx
dL}\approx
\frac{\alpha_{S}(4-4x+2x^{2})}{3\pi}\sqrt{\frac{2nC_{3}(x)}
{E_{q}x^{3}(1 -
x)}}\,.
\label{eq:3.170}
\eeq
The value of $\rho_{eff}$ in Eq. (\ref{eq:3.121}) can be
obtained
from the diffusion
relation
$\rho_{eff}\sim
(2\Delta\xi_{eff}/\mu_{q'g})^{1/2}$.
Here, as for the photon
radiation,
$\Delta\xi_{eff}\sim
L_{f}^{'}={\mbox{min}}(L_{f},1/|\Omega|)$.
This gives to a logarithmic
accuracy
$
\rho_{eff}\sim [ \alpha_{S}^{2} n
E_{q}x(1-x)]^{-1/4}\,.
$
The elimination of the infrared divergence is a
direct
consequence of the medium modification of the gluon formation
length.
At $\eta>>1$ the medium-modified formation
length
$L_{f}^{'}=L_{f}/\eta<<L_{f}$, and the
typical
transverse
size of virtual $q'g$ system becomes small
$\rho_{eff}<<1/m_{g}$.
In this region the dynamics
is
scaling.
As a result,
the
radiation rate
(\ref{eq:3.170})
has only
a logarithmic
dependence on the gluon mass coming from the factor
$C_{3}$.
Using the double gluon formula for the dipole cross cross
section, 
we find from Eq. (\ref{eq:3.170}) at $x<<1$ for
QGP
\beq \frac{d P_{g}}{dx
dL}\approx
4\alpha_{S}^{2}\sqrt{\frac{
nC_{T}\log(2/\mu_{D}\rho_{eff})}
{3\pi
E_{q}x^{3}}}\,,
\label{eq:3.180}
\eeq
where $C_{T}$ is the second order Casimir invariant for the
color
center. For nuclear matter, after expressing $C_{3}$ through
gluon
density, Eq. (\ref{eq:3.170})
yields
\beq
\frac{d P_{g}}{dx
dL}\approx
4\alpha_{S}
\sqrt{\frac{
n\alpha_{s}
x_{B}g(x_{B},10/\rho_{eff}^{2})}
{6
E_{q}x^{3}}}\,.
\label{eq:3.190}
\eeq
Note that Eqs. (\ref{eq:3.180}) and (\ref{eq:3.190}) differ
from
predictions
of Refs. \cite{B3} and \cite{Levin} by the factors $\sqrt{2}/3$
and
$12$,
respectively.

Ignoring the contributions to the energy loss from the
two
narrow regions near $x\approx 0$ and $x\approx 1$, in
which
Eq. (\ref{eq:3.170}) is not valid, we find that, in the limit
of strong LPM effect,
the energy
loss
per unit length is
\beq
\frac{d\Delta
E_{q}}{dL}\approx
1.1 \alpha_{s}\sqrt{nC_{3}(0)
E_{q}}\,.
\label{eq:3.200}
\eeq

%-----------------------------------------------------------------
\subsection{ Quark energy loss in hadron-nucleus collisions}
%-----------------------------------------------------------------
Let us now consider induced gluon radiation of a
fast
quark incident on a slab of nuclear matter of
thickness
$L$. This situation simulates gluon
emission
in
hadron-nucleus
collisions.
From
Eq. (\ref{eq:3.20}) using the oscillator Green's
function
(\ref{eq:2.330})
after some algebra the radiation rate can be represented in the
form
\beq
\frac{dP_{g}}{dx}=
L n
\frac{d\sigma^{BH}_{}}{dx}S_{}(\eta,l)
\,,
\label{eq:3.210}
\eeq
where
\beq
l=L/L_{f}=\frac{L[m_{q}^{2}x^{2}+m_{g}^{2}(1-x)]}{2E_{q}x(1-x)}\,,
\label{eq:3.220}
\eeq
and $\eta$ is defined by
Eq. (\ref{eq:3.160}).
In terms of the
dimensionless
variables $\eta$ and $l$, the suppression factor $S(\eta,l)$
is given
by
\beq
S_{}(\eta,l)=S_{}^{(1)}(\eta,l)+2S_{}^{(2)}(\eta,l)+
S^{(3)}(\eta,l)
\,,
\label{eq:3.230}
\eeq
\beq
S_{}^{(1)}(\eta,l)=
\frac{3}{l\eta^{2}}
\mbox{Re}\,
\int\limits_{0}^{l\eta}dy_{1}
\int\limits_{0}^{y_{1}}dy_{2}
\exp\left(-\frac{iy_{2}}{\eta}\right)
\left\{\frac{1}{y_{2}^{2}}-
\left[\frac{\phi}{\sin(\phi
y_{2})}\right]^{2}\right\}\,,
\label{eq:3.240}
\eeq
\bea
S^{(2)}(\eta,l)=
\frac{3}{l\eta^{2}}
\mbox{Re}\,
\int\limits_{0}^{l\eta}dy_{1}
\int\limits_{0}^{\infty}dy_{2}
\exp\left[-\frac{i(y_{1}+y_{2})}{\eta}\right]
\nonumber\\
\times
\left\{\frac{1}{(y_{1}+y_{2})^{2}}-
\left[\frac{\phi}
{\cos(\phi y_{1})\left(\tan(\phi
y_{1})+\phi
y_{2}\right)}\right]^{2}\right\}\,,
\label{eq:3.250}
\eea
\bea
S^{(3)}(\eta,l)=
\frac{3}{l\eta^{2}}
\mbox{Re}\,
\int\limits_{0}^{\infty}dy_{1}
\int\limits_{0}^{\infty}dy_{2}
\exp\left[-\frac{i(y_{1}+y_{2}+l\eta)}{\eta}\right]
\nonumber\\
\times
\left\{\frac{1}{(y_{1}+y_{2}+l\eta)^{2}}-
\left[\frac{\phi}
{\phi (y_{1}+y_{2})\cos{(\phi
l\eta)}+(1-\phi^{2}y_{1}y_{2})
\sin{(\phi
l\eta)}}
\right]^{2}\right\}\,,
\label{eq:3.260}
\eea
with
$\phi=\Omega/|\Omega|=\exp(-i\pi/4)$.
The first term on the right-hand side of
(\ref{eq:3.230})
corresponds,
in Eq. (\ref{eq:3.20}), to the contribution from the
integration
region $\xi_{1}<\xi_{2}<L$. The second term is associated with the
region
$\xi_{1}<0<\xi_{2}<L$, which gives the same contribution as the
region
$0<\xi_{1}<L<\xi_{2}$.
The last
term
in (\ref{eq:3.230}) comes from the region $\xi_{1}<0$ and
$\xi_{2}>L$.
The variables in (\ref{eq:3.240}),
(\ref{eq:3.250}),
(\ref{eq:3.260}) in terms of those
in
(\ref{eq:3.20}) are as
follows:
$y_{1}=(L-\xi_{1})|\Omega|$,
$y_{2}=(\xi_{2}-\xi_{1})|\Omega|$ in
(\ref{eq:3.240}),
$y_{1}=(L-\xi_{1})|\Omega|$,
$y_{2}=(\xi_{2}-L)|\Omega|$ in
(\ref{eq:3.250}),
$y_{1}=-\xi_{1}|\Omega|$,
$y_{2}=(\xi_{2}-L)|\Omega|$ in
(\ref{eq:3.260}).
In deriving (\ref{eq:3.240}), (\ref{eq:3.250}), (\ref{eq:3.260})
we
have used a representation of the first Green's
function
in the square brackets in (\ref{eq:3.20}) in terms of
a
convolution of the oscillator and the vacuum Green's
functions.
At $L\rightarrow \infty$ the factors $S^{(2)}$ and $S^{(3)}$ in
Eq.
(\ref{eq:3.230}) vanish, while $S^{(1)}$ tends to the
infinite
medium suppression factor
(\ref{eq:2.360}).
The finite-size effects come into play at $L\lsim
L_{f}^{'}$.
For $L<<L_{f}^{'}$ Eqs. (\ref{eq:3.240}),
(\ref{eq:3.250}),
(\ref{eq:3.260})
yield  $S^{(1,2)}(\eta,l)\propto l^{2}$ and $S^{(3)}\approx
1$.

In numerical calculations we take $m_{g}=0.75$ GeV. This
value
of $m_{g}$ was obtained in Ref. \cite{BFKL1} from the
analysis
of HERA data on structure function $F_{2}$
within
the dipole approach \cite{BFKL2} to the BFKL
equation. It
is also
consistent
with the nonperturbative estimate \cite{Shuryak2} of the
gluon
correlation radius in QCD vacuum $R_{c}\approx 0.27$
fm.
Note that the hadronic size is bigger
than
$1/m_{g}\approx R_{c}$ by a factor $\sim
4-5$.
It is this circumstance that allows us to neglect the
interference
effects connected with gluon emission from different
quarks.
For real nuclei LPM suppression turns out to be
relatively
small. For this reason in Eq. (\ref{eq:3.121}) we
take
$C_{2}(\rho_{eff})=C_{2}(1/m_{g})$.
For scattering of the $q\bar{q}g$ system on a nucleon, we
find
from the double gluon model \cite{DGM} $C_{2}(1/m_{g})\sim 1.3-4$
where
the lower and upper bounds correspond to
the
$t$-channel
gluon propagators with mass 0.75 and 0.2 GeV,
respectively.
The latter choice allows one
to
reproduce the
dipole
cross section extracted from the data on vector
meson
electroproduction \cite{NNPZ}. However, there is every
indication
\cite{BFKL1,BFKL2}
that a considerable part of the dipole cross section
obtained
in \cite{NNPZ} comes from the nonperturbative effects for
which
our approach is not justified. For this reason we
take
$C_{2}(1/m_{g})=2$, which seems to
be
a plausible estimate for the perturbative component of
the
dipole cross section
\cite{BFKL1}.
For quark mass, which controls the transverse size of
the
$q\bar{q}g$
system at $x\approx 1$, we take $m_{q}=0.2$ GeV. Notice
that
our predictions for $\Delta E_{q}$ are insensitive to the value
of
$m_{q}$.

We performed calculations
taking
$n=0.15$ fm$^{-3}$ and
$\alpha_{s}=1/2$.
Our numerical results in the region $L\lsim 10$
fm
can be parametrized in the
form
$\Delta E_{q}\approx 0.1
E_{q}(L/10\,\mbox{fm})^{\beta}$
with $\beta\approx 0.9-1$ for $E_{q}\lsim 50$ GeV
and
$\beta\approx 0.85-0.9$ for $E_{q}\gsim 200$
GeV.
Our estimate is in a
good
agreement with the longitudinal energy flow
measured
in hard $pA$ collisions with dijet final state
\cite{E609}
and the energy loss obtained from the analysis of
the
inclusive hadron spectra in $hA$ interactions
\cite{QK}.
Note that our result differs drastically from
the prediction
by Brodsky and Hoyer \cite{Brodsky}: $\Delta E_{q}\approx
0.25
(L/1\, \mbox{fm})$
GeV.

Our numerical calculations give the energy and $L$-dependence
of
$\Delta E_{q}$ close to those for the Bethe-Heitler
regime.
This can be readily
understood
at a qualitative
level. Indeed,
for a quark incident on a target the radiation rate
can
be represented in the form analogous to Eq. (\ref{eq:2.410}) in
QED.
In the case of interest absorption effects at
the
longitudinal scale about the nucleus size play a
marginal
role due to small
transverse
size of the $q\bar{q}g$ system ($\sim 1/m_{g}$). As a result,
the
radiation rate must be close to the Bethe-Heitler one,
and
we immediately
obtain
$\Delta E_{q}\sim E_{q} L n
\alpha_{s}C_{3}(0)/m_{g}^{2}$.
Thus, we see that for real nucleus LPM
suppression
does not play an important role. Note that this clearly demonstrates
that the
approach that was used in Ref. \cite{Levin} and which assumes strong
LPM
suppression is not applicable to hadron-nucleus
collision.

%----------------------------------------------------------------
\subsection{ Energy loss of a quark produced inside a medium }
%----------------------------------------------------------------
For a quark produced inside a medium
the
probability of gluon emission  can also be written in
the
form (\ref{eq:3.210}). The suppression factor in this case is
given
by
\beq
S_{}(\eta,l)=S_{}^{(1)}(\eta,l)+S_{}^{(2)}(\eta,l)
\,,
\label{eq:3.280}
\eeq
where $S^{(1,2)}$ are defined by
Eqs. (\ref{eq:3.240}),
(\ref{eq:3.250}).
From Eqs. (\ref{eq:3.240}), (\ref{eq:3.250}) one can
obtain
$S(\eta,l)\approx -l^{2}\log{l}$ for $l<< 1$
.
The physical mechanism behind this suppression of
radiation
at small
$L$
is obvious: the energetic quark produced through a hard mechanism
loses
the soft component of its gluon cloud and radiation at
distances
shorter than
the
time required for regeneration of the
quark
gluon field turns out to be
suppressed.
Notice that a similar suppression of
photon
radiation from an electron after a hard Coulomb scattering
was
discussed long ago by Feinberg
\cite{Feinb}.

%---------------------------------------------------------------------
Before presenting the numerical results, let us
consider
the energy loss at a qualitative
level.
We begin with the case of a sufficiently large
$E_{q}$
such that the maximum value
of
$L_{f}^{'}$, $L_{f}^{'}(\mbox{max})$, is much bigger than
$L$.
Taking into account the finite-size suppression
of
radiation at $L_{f}^{'}\gsim L$, we find that at high
quark
energy $\Delta
E_{q}$
is dominated by the contribution from two narrow regions of
$x$:
\beq
x\lsim \delta_{g}\approx L
m_{g}^{2}/2l_{0}E_{q}\,,~~~~~~~~
(1-x)\lsim \delta_{q}\approx L
m_{q}^{2}/2l_{0}E_{q}\,,
\label{eq:3.281}
\eeq
where
$l_{0}=\mbox{min}(1,1/\eta)$.
In both the regions the finite-size effects are
marginal
and the energy loss can be estimated using the infinite
medium
suppression factor. For
instance,
\beq
\Delta E_{q}(x\lsim
\delta_{g})\sim
\frac{16\alpha_{s}C_{3}(0)E_{q}Ln}{9\pi
m_{g}^{2}}
\int\limits_{0}^{\delta_{g}}dxS(\eta(x),l=\infty)\,.
\label{eq:3.290}
\eeq
Using Eq. (\ref{eq:3.160}) one can show that
$\eta(x\lsim
\delta_{g})\lsim
1$ at $L\lsim m_{g}^{2}/2nC_{3}(0)$. In this region
of
$L$ in (\ref{eq:3.290}) we can put
$S(\eta(x),l=\infty)\approx
1$ and
find
\beq
\Delta E_{q}\sim 0.25 \alpha_{s}C_{3}(0)n
L^{2}\,.
\label{eq:3.300}
\eeq
%which does not depend on the quark energy and gluon mass.
At $L\gg m_{g}^{2}/2nC_{3}(0)$ the typical values of
$\eta$
in (\ref{eq:3.290}) are much bigger than unity, and
using
the asymptotic
formula
for the suppression factor we
obtain
\beq
\Delta E_{q}\sim \alpha_{s}C_{3}(0)n
L^{2}\,.
\label{eq:3.310}
\eeq
A similar analysis for $x$ close to
unity
gives the contribution to $\Delta E_{q}$ suppressed
by
the factor $\sim 1/4$ as compared to that for small
$x$.
Thus we see that at high energy $\Delta
E_{q}$
does not depend on quark energy,
and
despite
the $1/m_{g,q}^{2}$ infrared divergence of the
Bethe-Heitler
cross section has only a
smooth
$m_{g}$-dependence originating from the factor
$C_{3}$.
We emphasize that the above analysis of the origin of
the
leading contributions makes it
evident
that $L^{2}$
dependence
of $\Delta E_{q}$ cannot be regarded as a direct consequence
of
LPM suppression of the radiation
rate
due to small angle multiple
scattering.

The finite-size effects can be
neglected
and $\Delta E_{q}$ becomes proportional to
$L$
if
$L_{f}^{'}(\mbox{max})\ll
L$.
If in addition the typical values of $\eta$ are much bigger than
unity,
then the energy loss per unit length is given by
formula
(\ref{eq:3.200}).

%-----------------------------------------------------
To study the infrared sensitivity of $\Delta E_{q}$,
we
performed numerical calculations for two values of
mass
of the radiated gluon $m_{g}=0.75$ and $m_{g}=0.375$
GeV.
As in the case of a quark incident on a nucleus, we
take
$m_{q}=0.2$ GeV and
$C_{2}(\rho_{eff})=C_{2}(1/m_{g})$.
In the case of QGP we take $T=250$ MeV, and
$\alpha=1/3$.
For scattering of the $q\bar{q}g$ system on a quark and a
gluon
we use for
$C_{2}(1/m_{g})$
predictions of the double gluon formula with
the
Debye screened gluon
exchanges.
In the region $L\lsim 5$ fm our numerical results can be
parametrized
in the
form
\beq
\Delta E_{q}\approx
D\left(\frac{L}{5\,\mbox{fm}}\right)^{\beta}\,.
\label{eq:3.320}
\eeq
The
$D$ and $\beta$ as function of $E_{q}$ are
shown
in Fig.~4 (nuclear matter) and Fig.~5 (QGP). In the
region
$5\lsim L\lsim 10$ fm $\beta$ in (\ref{eq:3.320}) is by 10-20
\%
smaller than for $L\lsim 5$
fm.
Note that $L_{f}^{'}(\mbox{max})\sim 5-10$ fm
for
$E_{q}\sim 10-40$ GeV in the case of nuclear
matter,
and $E_{q}\sim 150-600$ GeV for QGP. Then from Figs.~4,~5 one
can
conclude that the onset of the $L^{2}$ regime occurs
at
$L_{f}^{'}(\mbox{max})/L\gsim 2$. The closeness of $\beta$
to
unity at $E_{q}\approx 10$ GeV for QGP agrees with a small
value
of $L_{f}^{'}(\mbox{max})$ ($\sim 1$ fm). Our results
show
that the $m_{g}$-dependence of $\Delta E_{q}$ becomes weak
at
$E_{q}\gsim 50$ GeV. However, it is sizeable for $E_{q}\sim
10-20$
GeV.

Our predictions for $\Delta E_{q}$ must be regarded
as
rough estimates with uncertainties of at least a factor of 2 in
either
direction.
Nonetheless, rather large values of $\Delta E_{q}$
obtained
for QGP indicate that the jet
quenching
may be an important potential probe
for
formation of the deconfinement phase in $AA$
collisions.
A small quark energy loss obtained for nuclear matter
indicates
that the extraction of $\Delta E_{q}$ from experimental
data
on deep inelastic scattering on nuclei is a delicate
problem.

%---------------------------------------------------------------------
\section{LPM suppression in hard reactions on nuclear targets}
%---------------------------------------------------------------------
Another important example of the LPM effect in QCD is the
well-known shadowing in hard reactions on nuclear
targets.
For instance, nuclear shadowing in deep inelastic
scattering
at small values of the Bjorken
variable
$x_{B}=Q^{2}/2E_{\gamma^{*}}m_{p}$,
here $Q^{2}$ and $E_{\gamma^{*}}$ are the photon
virtuality
and energy,
respectively.
This effect is similar to LPM
suppression
of pair production in QED. Calculation of the valence
$q\bar{q}$
component of the shadowing
correction
$\Delta\sigma(\gamma^{*}A)=\sigma(\gamma^{*}A)-A\sigma(\gamma^{*}N)$
to $\gamma^{*}A$ total cross section in
the
limit $x_{B}\rightarrow 0$ can be
performed
within the frozen-size approximation
\cite{NZzp49}.
The light-cone path integral formalism allows one
to
take into account the parton transverse motion effects,
which
are important for evaluation of $x_{B}$-dependence of
nuclear
shadowing. Nonetheless, an accurate analysis, requiring evaluation
of
medium effects for the higher $q\bar{q}g_{1}...g_{n}$ Fock states,
is
a difficult problem. However, within the Double-Leading-Log
Approximation
(DLLA)
calculation of the leading
twist
contribution to $\Delta\sigma(\gamma^{*}A)$ is greatly
simplified.

In the
DLLA
the parton light-cone variables and the transverse
separations
for the $q\bar{q}g_{1}...g_{n}$ state are
ordered
\beq
x_{B}<<x_{n}<<x_{n-1}<<\ldots<<x_{1}<<x<1\,\,,
\label{eq:3.330}
\eeq
\beq
\frac{1}{Q^{2}}<<\rho^{2}<<\rho_{1}^{2}<<\ldots<<\rho_{n}^{2}\lsim
\frac{1}{m_{g}^{2}}\,.
\label{eq:3.340}
\eeq
As a result, in calculating the leading twist shadowing
correction,
the subsystem
$q\bar{q}g_{1}...g_{n-1}$
can be treated as a pointlike color-octet particle. Due to
the
ordering
in
the light-cone fractional momentum (\ref{eq:3.330})
the
transverse
motion of the center-of-mass of the
$q\bar{q}g_{1}...g_{n-1}$
subsystem can be neglected, and
only
the motion of the softest gluon ($g_{n}$) must be taken into
account.
Consequently, we can write $\Delta\sigma(\gamma^{*}A)$
as
\beq
\Delta \sigma(\gamma^{*}A)=\Delta
\sigma_{val}(\gamma^{*}A)
+\Delta\sigma_{3\Pom}(\gamma^{*}A)\,,
\label{eq:3.350}
\eeq
where
$\Delta\sigma_{val}(\gamma^{*}A)$
corresponds to the $q\bar{q}$ Fock state of the
virtual
photon,
while $\Delta\sigma_{3\Pom}(\gamma^{*}A)$ gives the
contribution
associated with the higher $q\bar{q}g_{1}...g_{n}$ Fock
states
treated as a two-body octet-octet
state.
Both the terms on the right-hand side of (\ref{eq:3.350}) can
be
written in the form similar to
Eq. (\ref{eq:2.470}).
For $\Delta\sigma_{val}(\gamma^{*}A)$ one can
obtain
(to simplify notation we do not indicate spin
variables)
\bea
\Delta\sigma_{val}(\gamma^{*}A)=
-\frac{1}{2}\mbox{Re}
\sum\limits_{q}
\int\limits_{0}^{1}
dx
\int d
{\bb}
\int\limits_{-\infty}^{\infty}
dz_{1}n(\bb,z_{1})
\int\limits_{z_{1}}^{\infty}
dz_{2}n(\bb,z_{2})
\int
d\r\,
\Psi_{\gamma^{*}}^{q*}(x,\r)\nonumber\\
\times\sigma_{2}(\rho
)
\Phi_{\gamma^{*}}^{q}(x,\r,\bb,z_{1},z_{2})
\exp\left[-\frac{i(z_{2}-z_{1})}{L_{f}^{q}}\right]\,,
\label{eq:3.360}
\eea
where
\beq
L_{f}^{q}=\frac{2E_{\gamma^{*}}x(1-x)}{m_{q}^{2}+Q^{2}x(1-x)}\,,
\eeq
\beq
\Phi_{\gamma^{*}}^{q}(x,\r,\bb,z_{1},z_{2})=
\int d\r'{\cal
K}_{q}(\r,z_{2}|\r',z_{1})
\Psi_{\gamma^{*}}^{q}(x,\r')\,\sigma_{2}(\rho')\,,
\label{eq:3.370}
\eeq
$n(\bb,z)$ is the nuclear density, $Q^{2}$ is the
photon
virtuality, $\Psi_{\gamma^{*}}^{q}(x,\r)$ is the light-cone
wave
function for transition $\gamma^{*}\rightarrow
q\bar{q}$.
In
Eq. (\ref{eq:3.370})
${\cal
K}_{q}$
is the Green's function for the
Hamiltonian
\beq
{\cal
H(\bb)}=\frac{{\qb}^{2}}{2\mu_{q\bar{q}}}+v(\bb,\r,z)\,,
\label{eq:3.380}
\eeq
where
\beq
v(\bb,\r,z)=-i\frac{n(\bb,z)\sigma_{2}(|\r|)}{2}\,,
\label{eq:3.390}
\eeq
and
$\mu_{q\bar{q}}=E_{\gamma^{*}}x(1-x)$.

Using the light-cone wave functions for
the
$q\bar{q}g_{1}...g_{n}$ Fock
states
obtained in
Ref.
\cite{NZJETP78},
we can represent
$\Delta\sigma_{3\Pom}(\gamma^{*}A)$
in the
form
\bea
\Delta\sigma_{3\Pom}(\gamma^{*}A)=
-\frac{1}{2}\left(\frac{9}{4}\right)^{2}
\mbox{Re}
\int\limits_{x_{B}}^{1}
dx_{g}
\int d
\bb
\int\limits_{-\infty}^{\infty}
dz_{1}n(\bb,z_{1})
\int\limits_{z_{1}}^{\infty}
dz_{2}n(\bb,z_{2})
\int d\r\, \lambda^{2}(x_{g},Q^{2})
\nonumber\\
\times
\Psi_{\gamma^{*}}^{g*}(x_{g},\r)
\sigma_{2}(\rho
)
\Phi_{\gamma^{*}}^{g}(x_{g},\r,\bb,z_{1},z_{2})
\exp\left[-\frac{i(z_{2}-z_{1})}{L_{f}^{g}}\right]\,,~~~~~
\label{eq:3.400}
\eea
where
$$
L_{f}^{g}=\frac{2E_{\gamma^{*}}x_{g}}{m_{g}^{2}}\,,
$$
\beq
\Phi_{\gamma^{*}}^{g}(x_{g},\r,\bb,z_{1},z_{2})=
\int d\r'{\cal
K}_{g}(\r,z_{2}|\r',z_{1})
\Psi_{\gamma^{*}}^{g}(x_{g},\r')\,\sigma_{2}(\rho')\,.
\label{eq:3.410}
\eeq
The Hamiltonian for the Green's
function
${\cal K}_{g}(\r,z_{2}|\r',z_{1})$ in Eq. (\ref{eq:3.410}) can
be
obtained from Eqs. (\ref{eq:3.380}),
(\ref{eq:3.390})
replacing $\mu_{q\bar{q}}$ by
$\mu_{q\bar{q}g}=E_{\gamma^{*}}x_{g}$,
and $\sigma_{2}$ by $\frac{9}{4}\sigma_{2}$. The
factor
$(9/4)^{2}$ in Eq. (\ref{eq:3.400}) reflects the fact
that
the dipole cross section for octet-octet state
equals
$(9/4)\sigma_{2}(\rho)$. The factor $\lambda^{2}$
in
Eq. (\ref{eq:3.400})
coming
from the internal $q\bar{q}g_{1}\ldots g_{n-1}$ states is given
by
\beq
\lambda^{2}(x_{g},Q^{2})=
\frac{4I_{0}(2\sqrt{\zeta})}{3\pi^{2}}
\sum\limits_{q}
\int\limits_{0}^{1}dx\int d\r
\rho^{2}\alpha_{s}(\rho)
|\Psi_{\gamma^{*}}^{q}(x,\r)|^{2}
\,,
\label{eq:3.420}
\eeq
where
$$
\zeta=\frac{12}{\beta_{0}}
\log{\left(\frac{\alpha_{s}(m_{g}^{2})}{\alpha_{s}(Q^{2})}\right)}
\log{\left(\frac{1}{x_{g}}\right)}
$$
is the expansion parameter of the DLLA,
and
$I_{0}(z)\approx \exp(z)/\sqrt{2\pi z}$ is
the
Bessel
function.
The
light-cone
wave function describing the softest
gluon
entering Eqs. (\ref{eq:3.400}), (\ref{eq:3.410}) is given
by
\cite{NZJETP78}
$$
\Psi_{\gamma^{*}}^{g}(x_{g},\r)=
\frac{m_{g}}{r\sqrt{x_{g}}}\left[
K_{1}(m_{g}\rho_{1})\frac{\eb^{*}\r_{1}}{|\r_{1}|}-
K_{1}(m_{g}\rho_{2})\frac{\eb^{*}\r_{2}}{|\r_{2}|}\right]\,,
$$
where $\eb$ is the gluon polarization
vector,
and $\r_{1,2}=\r \pm\rb/2$, $|\rb|\sim
1/Q$.
Note that, according to the derivation of
Eqs. (\ref{eq:3.360}),
(\ref{eq:3.400}), the perturbative component of the dipole
cross
section
entering these equations, must be evaluated in the
Born
approximation.

Similar expressions can be obtained for shadowing
corrections
in Drell-Yan pair and heavy quark
production.
The results of numerical calculations of nuclear shadowing
in
hard reactions will be presented
elsewhere.

%-------------------------------------------------------------
\section{Conclusion}
%-------------------------------------------------------------
We have discussed a new
approach
to the LPM effect in QED and QCD. This approach is
based
on the path integral representation of the
light-cone
wave functions. Using the unitarity we
express
the cross section of the radiation
process
$a\rightarrow bc$ in terms of the radiative correction to
the
transverse propagator of particle
$a$.
Evaluation of the cross section of transition
$a\rightarrow
bc$ is reduced to solving the
two-dimensional
Schr\"odinger
equation with an imaginary potential proportional to the total
cross
section of interaction of $\bar{a}bc$ state with a
medium
constituent. We have demonstrated a close relationship between
LPM
suppression
for
the radiation process $a\rightarrow bc$ and the
absorption
correction for $\bar{a}bc$
state.

For bremsstrahlung in QED we have evaluated the LPM effect
for
finite-size homogeneous and structured
targets.
For structured
targets
we predict minima and maxima in the photon
spectra.
We have given a rigorous treatment of the
Coulomb
effects, which were previously treated only to
a logarithmic
accuracy.
We have also included the inelastic process neglected in
previous
works.
For the first time we have performed a rigorous theoretical
analysis
of
the experimental data on the LPM effect obtained at SLAC
\cite{SL1}.
The theoretical predictions are in very good agreement with the
spectrum
measured at SLAC \cite{SL1} for the
homogeneous
gold target with $L=0.7\%X_{0}$ for 25 GeV electron
beam.

For the first time we have performed a rigorous analysis
of
the induced gluon radiation in cold nuclear matter and
in
QGP within GW model
\cite{GW1}.
For a quark incident on a nucleus we
predict
$\Delta E_{q}\approx 0.1
E_{q}(L/10\,\mbox{fm})^{\beta}$,
with $\beta$ close to
unity.
For a sufficiently energetic quark produced inside a
medium
we find the radiative energy loss $\Delta E_{q}\propto
L^{2}$,
where $L$ is the distance passed by the quark in
the
medium. It has a weak dependence on the initial quark
energy.
The $L^{2}$ dependence turns to $L^{1}$ as the quark
energy
decreases.

We have also demonstrated that the developed theory of the LPM
effect
can be used for an accurate evaluation of the leading
twist
contribution to nuclear shadowing in hard reactions on
heavy
nuclei.

{\large
\bf
\begin{flushleft}
Acknowledgements
\end{flushleft}
}

I would like to
thank
R.Baier, Yu.L.Dokshitzer, P.Hoyer, J.Knoll, A.H.Mueller,
N.N.Nikolaev,
S.Peigne and D.Schiff for
discussions.
This work was partially supported by the INTAS
grants
93-239ext and
96-0597.

\newpage
\begin{center}
{\Large \bf Figures}
\end{center}

\begin{figure}[h]
\begin{center}
\epsfig{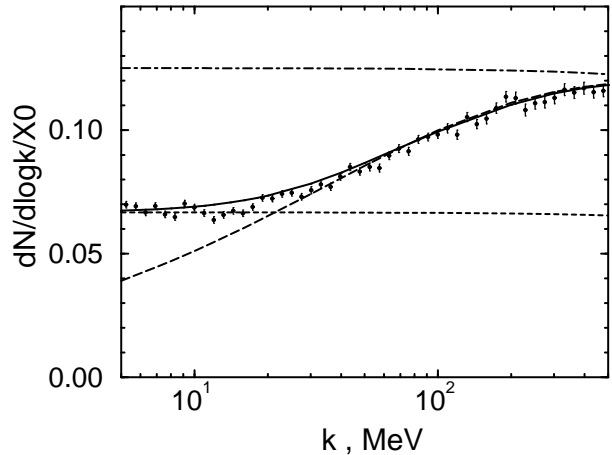}
\end{center}
\caption{The bremsstrahlung spectrum for 25 GeV electrons
incident
on a gold target with a thickness of
$0.7\%X_{0}$.
The experimental data are from
Ref. [22]
%\cite{SL1}.
The full curve shows our results obtained
using
%Eqs.~(\ref{eq:2.410}),~(\ref{eq:2.460}),~(\ref{eq:2.470}). The
Eqs.~(46),~(51),~(52). The
dashed
curve
was
obtained in the frozen-size approximation (56)
%(\ref{eq:2.510}).
The long-dashed curve shows the spectrum for the
infinite
medium. The Bethe-Heitler spectrum is shown by the dot-dashed
curve.}
\label{f1}
\end{figure}

\begin{figure}[h]
\begin{center}\epsfig{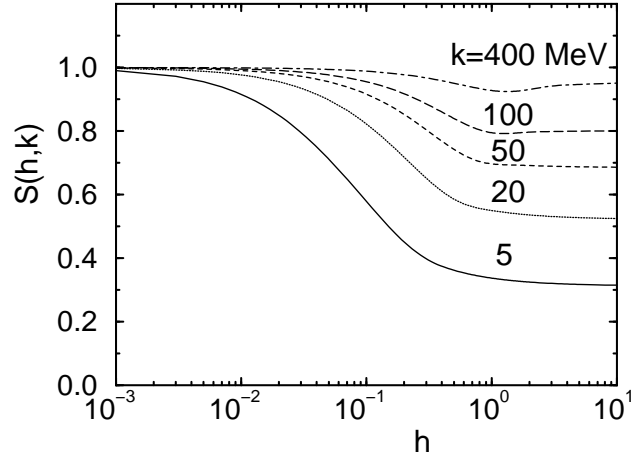}\end{center}
\caption{
The LPM suppression factor for 25 GeV electron
incident
on a homogeneous gold target as a function of the
ratio
$h=L/L_{f}$ and the photon
momentum.
}
\label{f2}
\end{figure}
\begin{figure}[h]
\begin{center}\epsfig{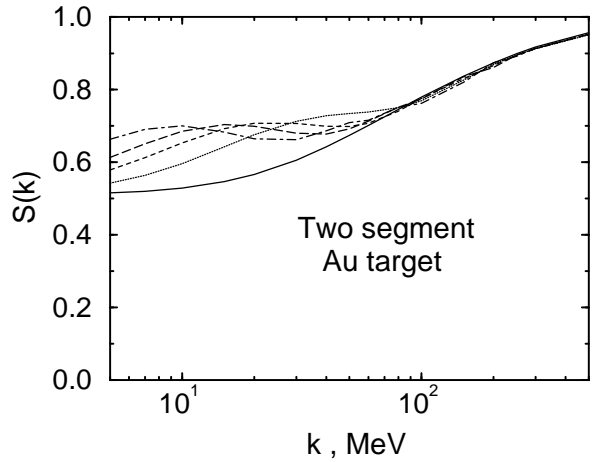}\end{center}
\caption{
The LPM suppression factor for 25 GeV electron incident
on
a two segment gold target. The thickness of each segment is
$0.35\%X_{0}$.
The set of gaps is as follows: 0 (solid
curve),
$0.7\%X_{0}$ (dotted curve), $1.4\%X_{0}$ (dashed
curve),
$2.1\%X_{0}$ (long-dashed curve), $3.5\%X_{0}$ (dot-dashed
curve).
}
\label{f3}
\end{figure}

\begin{figure}[h]
\begin{center}\epsfig{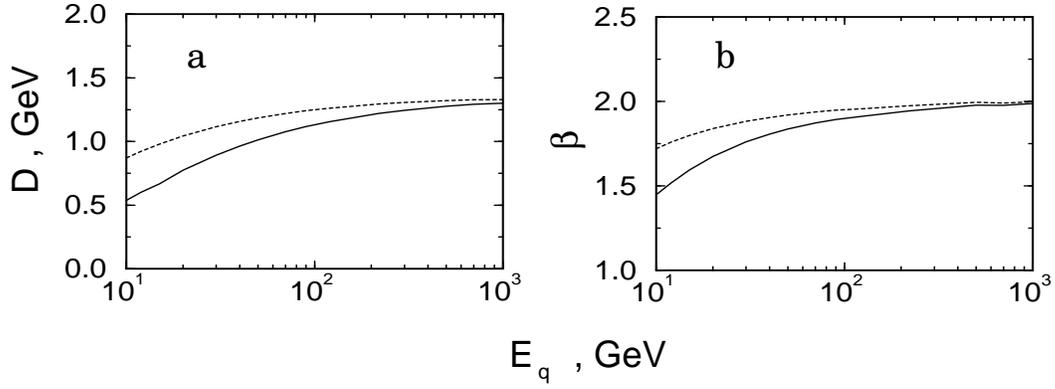}\end{center}
\caption{
The parameters $D$ (a) and $\beta$ (b) for the
parametrization
%(\ref{eq:3.320}) for nuclear matter. The solid lines
(96) for nuclear matter. The solid lines
correspond
to $m_{g}=0.75$ GeV, and the dashed ones to $m_{g}=0.375$
GeV.
}
\label{f4}
\end{figure}

\begin{figure}[h]
\begin{center}\epsfig{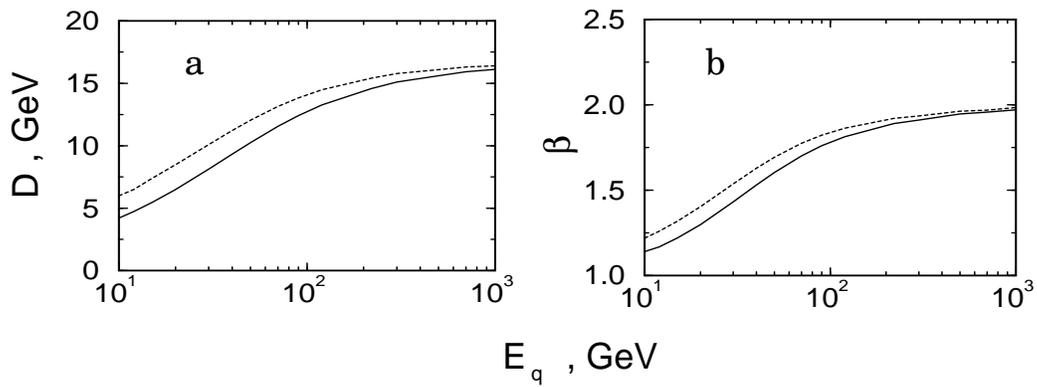}\end{center}
\caption{
The same as in Fig.~4 but for QGP at $T=250$
MeV.
}
\label{f5}
\end{figure}

\end{document}